\documentclass{aa}  

\usepackage{graphicx}
\usepackage{txfonts}
\usepackage{natbib}
\usepackage{longtable}
\bibpunct[; ]{(}{)}{;}{a}{}{,}

\begin{document}

   \title{Orbital and physical parameters of eclipsing binaries from the All-Sky Automated Survey catalogue}

   \subtitle{X. Three high-contrast systems with secondaries detected with IR spectroscopy}

\author{K. G. He\l miniak \inst{1}
	\and
	A. Tokovinin\inst{2}
	\and
	E. Niemczura\inst{3}
	\and
	R. Paw\l aszek\inst{1}
	\and
	K. Yanagisawa\inst{4}
	\and
	R. Brahm\inst{5}
	\and
	N. Espinoza\inst{6,5}
	\and
	N.~Ukita\inst{4,7}
	\and
	E. Kambe\inst{8}
	\and
	M. Ratajczak\inst{3}
	\and
	M. Hempel\inst{5}
	\and
	A.~Jord\'an\inst{9,5,6}
	\and
	M.~Konacki\inst{1}
	\and
	P.~Sybilski\inst{1}
	\and
	S. K. Koz\l owski\inst{1}
	\and
	M.~Litwicki\inst{1}
	\and
	M.~Tamura\inst{10,11,12}
          }

   \institute{
Nicolaus Copernicus Astronomical Center, Polish Academy of Sciences, ul. Rabia\'{n}ska 8, 87-100 Toru\'{n}, Poland\\ \email{xysiek@ncac.torun.pl}
	\and
Cerro Tololo Inter-American Observatory, Casilla 603, La Serena, Chile 
	\and
Astronomical Institute, University of Wroc{\l}aw, Kopernika 11, 51-622 Wroc{\l}aw, Poland
	\and
Okayama Astrophysical Observatory, National Astronomical Observatory of Japan, 3037-5 Honjo, Kamogata, Asakuchi,\\Okayama 719-0232, Japan 
	\and
Instituto de Astrof\'isica, Pontificia Universidad Cat\'{o}lica de Chile, Av. Vicu\~{n}a Mackenna 4860, 7820436 Macul, Santiago, Chile
	\and
Max-Planck-Institut f\"ur Astronomie, K\"onigstuhl 17, Heidelberg 69 117, D-69117, Germany
	\and
The Graduate University for Advanced Studies, 2-21-1 Osawa, Mitaka, Tokyo 181-8588, Japan
	\and
Subaru Telescope, National Astronomical Observatory of Japan, 650 North Aohoku Place, 
Hilo, HI 96720, USA
	\and
Millennium Institute of Astrophysics, Av.\ Vicu\~na Mackenna
4860, 7820436 Macul, Santiago, Chile
	\and
Department of Astronomy, The University of Tokyo, 7-3-1, Hongo, Bunkyo-ku, Tokyo, 113-0033, Japan
	\and
Astrobiology Center of NINS, 2-21-1, Osawa, Mitaka, Tokyo 181-8588, Japan 
	\and 
National Astronomical Observatory of Japan, 2-21-1 Osawa, Mitaka, Tokyo 181-8588, Japan 
             }

   \date{Received ...; accepted ...}

 
  \abstract
   {}
   {We present results of the combined photometric and spectroscopic analysis of three
detached eclipsing binaries, which secondary components are not visible or very 
hard to identify in the optical spectra -- ASAS~J052743-0359.7, ASAS~J065134-2211.5, and
ASAS~J073507-0905.7. The first one is a known visual binary ADS~4022, and we found that
it is a quadruple system, composed of two spectroscopic binaries, one of which shows
eclipses. None of the systems was previously recognized as a spectroscopic binary.}
   {Using the following telescopes/spectrographs: Subaru/IRCS, CTIO-1.5m/CHIRON, 
Euler/CORALIE, MPG-2.2m/FEROS, OAO-188/HIDES, and TNG/HARPS-N, we collected a number of 
high-resolution optical and IR spectra to calculate the radial velocities (RVs)
and later combined them with MITSuME and ASAS photometry. The Subaru/IRCS infra-red 
spectra were crucial for secure identification of the cooler components' lines. 
RV measurements were done with the TODCOR technique, and RV curves 
modelled with our own procedure V2FIT. Light curve modelling was performed
with JKTEBOP and PHOEBE codes. Temperatures and metallicities of two systems were 
estimated from spectra. For the ADS~4022 system we also used the archival WDS 
data and new SOAR observations in order to derive the orbit of the visual pair for the 
first time. Ages were estimated by comparing our results with PARSEC isochrones.}
   {The eclipsing pair ASAS~J052743-0359.7~A ($P=5.27$~d) is composed of a 
1.03(6)~M$_\odot$, 1.03(2)~R$_\odot$ primary and a 0.60(2)~M$_\odot$, 
0.59(2)~R$_\odot$ secondary. The components of the $P=21.57$~d non-eclipsing pair B
likely have masses in between the two eclipsing components, and both pairs are on 
a $\sim$188 yr orbit around their common centre of mass. The system ASAS~J065134-2211.5 
($P=8.22$~d) consists of a 0.956(12)~M$_\odot$, 0.997(4)~R$_\odot$ primary and a
0.674(5)~M$_\odot$, 0.690(7)~R$_\odot$ secondary. Finally, ASAS~J073507-0905.7
($P=1.45$~d), which consists of a 1.452(34)~M$_\odot$, 1.635(12)~R$_\odot$ 
primary and a 0.808(13)~M$_\odot$, 0.819(11)~R$_\odot$ secondary, is likely a 
pre-main-sequence system. In all cases secondary eclipses are total.}
   {}


   \maketitle
%

\section{Introduction}
During the past recent years we have seen an increasing interest in
studies of low mass, K and M-type dwarfs, driven mainly by the exoplanet
search surveys with dedicated instruments, both photometric 
\citep{bur17,whe17} and spectroscopic \citep{qui10,art14,kot14}. 
The goal behind such surveys is to find Earth-sized habitable planets, 
that are easier to detect around cool dwarfs, than around F or G-type
stars. However, behind each exoplanet detection and characterisation lays
the knowledge of the host star, and the K and M dwarfs, despite 
constituting about 90\% stars of the Galaxy, are relatively poorly studied.
For decades they have been known to show significant discrepancies between
the observed and predicted radii and temperatures \citep[e.g.][]{lac77,pop97,tor02,mor09a}.
Only recently the stellar structure and evolution models start to treat the
convection and activity on low-mass stars in a way that allows to reduce
these inconsistencies.

Reliable tests of these models can only be done with sufficiently precise
observational data, mainly masses and radii derived for components of detached
eclipsing binaries \citep{las02,tor10}. The number of systems with
the parameters measured well enough (relative errors below $2-3$\%) 
is constantly rising, but the vast majority of them are double-K or 
double-M-type pairs that are relatively faint, thus difficult to follow 
up spectroscopically. Only two such pairs brighter than $V=10$~mag -- 
\object{YY~Gem} \citep[$V=9.83$ in {\it Simbad};][]{tor02} and \object{1SWASP~J093010.78+533859.5} 
\citep[$V=9.84$;][]{koo14} -- can be found in the on-line catalogue DEBCat 
\citep{sou15}\footnote{\tt http://www.astro.keele.ac.uk/$\sim$jkt/debcat/}. 
Another one -- \object{AK~For} \citep[$V=9.13$;][]{hel14} -- has very good mass 
measurements, but slightly worse radius ($2.6-2.9$\%). However, three other 
pairs found in DEBCat -- \object{IM~Vir} \citep[$V=9.57$;][]{mor09b}, \object{V530~Ori}
\citep[$V=9.96$;][]{tor14} and \object{V2154~Cyg} \citep[$V=7.83$;][]{bri17} 
-- as well as another bright system \object{V1200~Cen} \citep[$V=8.50$;][]{cor15}, 
prove that low-mass dwarfs can be found and studied 
in bright binaries as companions to stars of earlier types. A large number
of pairs with secondaries of even lower mass than in V1200~Cen and V530~Ori,
has been reported by \citet{tri17} on the basis on photometric data from the 
Wide Angle Search for Planets \citep[WASP; ][]{pol06}
as the outcome of the Eclipsing Binaries with Low Mass (EBLM)
project, but they don't have their dynamical masses given.

Such pairs themselves are valuable for testing stellar evolution 
models. Due to their low mass ratio, components are located relatively far 
from each other on the HR diagram, and differ
in internal structure, level of activity, etc. Testing the models is thus
more demanding than in the case of nearly identical pairs.
Low-mass-ratio pairs are also important for the theories of 
binary stars formation. For example, different formation models can be
distinguished by different distributions of masses of secondaries, or mass
ratios in binaries \citep{bos93,bat00}. Unfortunately, the lower ends of these 
distributions are still poorly sampled due to observational restrictions --
cooler secondaries are many times fainter than the hotter primaries, which
makes them difficult do detect directly.

In our spectroscopic survey of southern detached eclipsing binaries (DEBs)
we have initially focused on searching for low-mass pairs, identifying
several new ones \citep{hel11a,hel11b,hel12,hel14}, but we have found
even more relatively bright, single-lined systems with F and G-type primaries 
and faint secondaries not seen or very difficult to identify in the 
optical spectra\footnote{It should be
noted that not all SB1 systems have unseen secondaries that are cooler, thus better 
visible in IR. In significant number of DEBs, like systems with an evolved giant
or some classical Algols, the primary actually has lower $T_{\rm eff}$, but is 
significantly larger than the secondary, and contributes more to the overall flux.}. 
We have thus started a small, low-effort program of
infra-red (IR) spectroscopic observations with the Subaru telescope, 
targeting only few of the most northern objects in
our sample. In this paper we present three examples of those important
objects, each of them showing unique, interesting characteristics. 
We also show that IR observations of low-mass stars paired with hotter 
primaries can be an efficient way to obtain important stellar parameters.


\section{Detection of faint secondary components} 
The vast majority of spectroscopic observations are done in the optical 
regime. Lets consider a binary composed of a 5700~K primary (solar analogue) 
and a 4000~K ($\sim$0.6~M$_\odot$) secondary, both on the main sequence. 
A theoretical 5~Gyr, solar metallicity isochrone predicts magnitude difference 
in $V$ of 3.53~mag, which translates into flux ratio (secondary over primary) of
$l_2/l_1\simeq0.04$. Meanwhile, this ratio rises to 0.1 (2.5~mag difference) 
in the $I$ band, to 0.16 (2.0~mag) in $J$, and 0.24 (1.55~mag)
in $K$. The secondary is, therefore, harder to detect on shorter wavelengths,
especially in spectra of relatively low signal-to-noise ratio (SNR).
This was the main motivation behind our IR observations.
Such situation can be seen in the aforementioned compilation of 118 
single-lined spectroscopic binaries with M-type companions by \citet{tri17}.
The authors present a table with predicted brightnesses of secondaries in 
$V,R,J$ bands, which can be compared to apparent magnitudes of the primaries.
In average, the flux ratio of secondary to primary (from magnitude 
difference) in $J$ is 19 times larger than in $V$, and exceeds 60$\times$ in
single cases.

%

The usage of IR spectroscopy to detect faint companions
has been known in literature for years \citep{maz02,maz03,pra02},
and is still in use \citep{pra18}. Apart from the higher flux ratio, 
its another advantage is that it does not require extensive, additional 
observations. To obtain masses of both components, one can base on 
information derived from the previous observations in optical, 
i.e. the orbital solution built for a single-lined spectroscopic 
binary (SB1). In such case only one, maximum two 
parameters are to be found: the secondary's velocity semi-amplitude $K_2$,
and (optionally) the difference between systemic velocities of two 
components $\gamma_2-\gamma_1$. Since the number of free parameters is 
low (1-2), only few additional observations are enough to find them.
For more details, see \citet{maz02}.

For a pair of G+K type stars ($T_{\rm eff,1}=5700$~K, $T_{\rm eff,2}=4000$~K),
assuming a Poissonian distribution of the photon noise (${\rm SNR} \sim \sqrt{\rm flux}$),
having the signal from the secondary ($l_2$) 10 times above the noise ($\sigma$) 
requires a spectrum of SNR$\simeq$260 in the $V$ band, but only 75 in $J$.
Even if we assume that IR spectrographs have in general lower efficiencies 
than the optical ones, we still need exposures few times longer in $V$
than in $J$ to reach the same $l_2/\sigma$. Adding new, high- or 
very-high-SNR optical observations (which can be impossible for stars
as faint as $V=11$~mag without an easy access to a large-aperture
telescope), is therefore generally more challenging than taking few 
spectra in IR. 

There are advanced methods to extract the information about the faint secondary
(its RVs or atmospheric parameters) from single composite spectra, or
time series thereof. The Two-Dimensional Cross-Correlation technique
\citep[TODCOR; ][]{zuc94} compares an observed spectrum of an SB2 with two
template spectra, and looks for the maximum of a 2-D cross-correlation 
function (CCF) in a $v_1/v_2$ plane, where $v_1$ and $v_2$ are the RVs 
of the primary and secondary component. It works
relatively well when the difference between $v_1$ and $v_2$ is small 
(lines of both may components overlap), and does not require high SNR.
With a proper selection of templates, RV measurements of a faint secondary
are possible (which was done for some of the spectra in this study). 
These possibilities are however limited, and a maximum of the CCF coming 
from the secondary can be easily confused with side-peaks, which arise from 
coincidence of lines in the observed spectrum with different lines in the template,
or simply the noise. The ``proper'' value of $v_2$ may not be known without 
additional information.

\citet{tka03} introduced a method of denoising the spectra on the basis of
modified least-squares deconvolution \citep[LSD;][]{don97,koc10}. 
They show significant improvement of the SNR, up to 10 times, on
single- and double-star spectra, which initial SNR was down to 35 
(for synthetic spectra) or 50 (for real observations). The 
undoubtful advantage is that the SNR is improved without loosing important
information (e.g. line depths and asymmetries), so an analysis to obtain 
atmospheric parameters can be run. The LSD has been successfully applied 
to measure the RVs of both stars in a large-contrast binary on 
multiple occasions, but specific exceptions are known in the literature
\citep[e.g. KIC~5640750 in ][]{the18}. Some prior knowledge, or good assumptions of the 
atmospheric parameters (i.e. $T_{\rm eff}, \log(g)$) is also required.


Finally, there is a variety of techniques that allow to obtain spectra of two
components separately. They base, for example, on disentangling in wavelength
domain \citep{sim94} or Fourier transform \citep{had95}, or tomographic
decomposition \citep{bag91,kon10}. They require multiple observations
that cover various orbital phases (different RV shifts), and spectra
of at least moderate SNR. They also work best on homogeneous samples, 
meaning spectra taken with the same instrument under the same setting. 
Since the resulting spectrum is obtained from a combination of $N$ spectra, 
the gain in its SNR is of the order of $\sqrt{N}$. This means that to really 
improve the SNR, $N$ must be quite large (tens of spectra). This, in practice, 
requires a significant amount of telescope time dedicated for observations of
a given target. Without sufficient number of quality observations, the 
disentangled spectrum of a faint secondary may still have its SNR too low
for the purpose of further analysis \citep{deb13,the18}. 

Different
approaches and different realisations of LSD or disentangling have their 
own advantages and limitations. For example, while some of these methods require 
prior knowledge of the RVs, or at least the orbital parameters, others may give 
the complete orbital solution without RVs as the input. There are cases that are 
limited to only two components, but other approaches have been shown to successfully 
decompose spectra of up to five stars in a system. Nonetheless, they are useful to 
obtain separate spectra of fainter components with improved SNR, which can be 
used for spectroscopic analysis, and were successfully used in multiple studies.

In our spectroscopic survey we observed nearly 300 southern DEBs,
using a variety of telescopes and spectrographs, mainly of a 
1-2 meter class. Brightness of our targets in $V$ varies from $\sim$8.5
to $\sim$11.5~mag, and even $>$12~mag in few cases \citep{hel11b}. 
The survey required a substantial amount of telescope time, and optimization
of the observing strategy for the most important immediate objective -- 
calculation of radial velocities and masses for a large number of targets. 
This can be effectively done with TODCOR on spectra of SNR as low as $\sim20$,
or even lower, when a stable instrument is used. Therefore, for many cases 
we did not attempt to obtain a large number of very-high-SNR spectra of 
specific targets, especially with the smaller telescopes. Moreover, for a given system, 
the spectra are usually collected with few spectrographs, so the data are 
inhomogeneous. This is why, for large contrast objects from our survey, 
like those described in this study, we believe that the IR observations, 
with the aid of TODCOR for RV calculations, is the safest and most convenient approach, 
as it is direct, robust, and the least telescope time consuming (although we consider 
implementation of the modified LSD for our research in the nearby future). The direct 
comparison of our approach with other, previously mentioned ones (LSD, spectral 
decomposition) is out of the scope of this paper. 

\section{Objects}

The three presented targets were monitored spectroscopically as a part of
our ongoing program of high-resolution observations of DEBs selected
from the ASAS Catalog of Variable Stars \citep[ACVS;][]{poj02}, which
is a product of the All-Sky Automated Survey (ASAS). All of them have
been discovered as eclipsing binaries by the ASAS, and none of them 
has been known as a spectroscopic binary till now. This is the first
study of these three systems.

ASAS~J052743-0359.7 (\object{HD~35883}, TYC~4751-298-1, WDS~J05278-0400AB, ADS~4022~AB, 
hereafter ASAS-052) is a $V=9.50$~mag system, with 0.3~mag deep primary
eclipse and the secondary minimum almost invisible in the ASAS data.
It is associated with an X-ray source 1RXS~J052742.1-035947 \citep{vog99}.
It has been known as a visual binary since 1902 \citep{ait03}. The Washington 
Double Star catalogue \citep[WDS;][]{mas01} currently gives 18 measurements of 
separation and position angle, some of them with components' magnitudes 
or their difference. The first 13 come from years 1902-1978 and were the only
known at the time of Subaru/IRCS observations. Further speckle interferometric
observations with SOAR were done for this study. 
\citet{kha09} gives a somewhat uncertain parallax of 21.4$\pm$19.0~mas, 
probably affected by the binarity of the object. There is no parallax 
given in the {\it Gaia} Data Release~2 \citep[GDR2;][]{gai16,gai18},
but the catalogue lists the effective temperature of $5321^{+103}_{-128}$~K.
We immediately noticed three sets of lines in the spectra, but only
after several spectroscopic observations we have realized that the
system is not triple but a quadruple. Only the strongest set of lines 
follows the photometric period ($P\simeq5.27$~d), and the other two
were undoubtedly coming from the same SB2 of an unknown period. We failed
to identify the second set of the DEB component before the infra-red 
observations.

ASAS~J065134-2211.5 (BD-22~1566, PPM~251128, \object{TYC~5962-2159-1}, hereafter ASAS-065) 
is the longest-period system in this study ($P\simeq8.22$~d), and the faintest 
($V=9.90$~mag). ASAS data show a $\sim$0.7~mag deep primary eclipse, and 
almost invisible secondary. In our first optical data the detection of the 
secondary was dubious, and we failed to find it in many spectra taken later. 
This system is also associated with an X-ray source 
-- \object{1RXS~J065133.6-221121}. Results of a spectroscopic analysis can be found in the 
5th data release of the RAdial Velocity Experiment catalogue \citep[RAVE;][]{kun17}.
The estimates of (calibrated) effective temperature (5331$\pm$57~K), gravity 
(4.43$\pm$0.08~dex), and metallicity (-0.10$\pm$0.09~dex), as well as 
spectrophotometric parallax (12.8$\pm$5.5~mas) are given. The parallax and
$T_{\rm eff}$ from GDR2 are formally in agreement (10.79$\pm$0.03~mas and 5175$^{+489}_{-92}$,
respectively).

ASAS~J073507-0905.7 (\object{HD~60637}, BD-08~1989, TYC~5397-1982-1, hereafter ASAS-073)
is the brightest star in our sample ($V=9.30$~mag), and has the shortest
orbital period ($P\simeq1.45$~d). It shows a 0.33~mag deep primary eclipse, and
0.05~mag deep secondary -- deepest in the sample. As expected from such
light curve, the secondary star was more prominent in the optical spectra than
in case of other presented system, but still we were not always able to detect
it. Moreover, it is the only system for which $I$-band ASAS light curve is
available. On the basis of multi-band photometry \citet{amm06} derived 
$T_\mathrm{eff}=6314^{+76}_{-84}$~K, [Fe/H$]=0.01^{+16}_{-17}$~dex, and 
estimated the distance to be $75^{+70}_{-28}$~pc. The GDR2 distance
is, however, 171.6$\pm$1.2~pc, but the effective temperature
is similar: $6323^{+233}_{-116}$~K.

\section{Observations and data reduction}
\subsection{Optical spectroscopy}
The optical spectra were taken with a number of spectrographs that we use
for our spectroscopic survey. The CHIRON spectrograph \citep{sch12,tok13}, 
attached to the 1.5-m telescope in CTIO (Chile), was used in the ``slicer'' mode, 
which provides spectral resolution of $R\sim90\,000$. This is the only telescope 
described here that works in service mode. Spectra were reduced with 
the pipeline developed at Yale University \citep{tok13}. Wavelength 
calibration is based on ThAr lamp exposures taken just before the science 
observation. Barycentric corrections are not applied by the pipeline, thus we
were calculating them ourselves under IRAF\footnote{IRAF is written and supported by the 
IRAF programming group at the National Optical Astronomy Observatories (NOAO) in Tucson, 
AZ. NOAO is operated by the Association of Universities for Research in Astronomy (AURA), 
Inc. under cooperative agreement with the National Science Foundation. 
http://iraf.noao.edu/} with {\it bcvcor} task. 
For the targets described here we did not use the ``fiber'' mode (more efficient 
but less precise than ``slicer''), nor the available iodine (I$_2$) cell. Without 
the I$_2$, the stability of the instrument is estimated to be better than 
15 m/s. For the radial velocity (RV) measurements we used 36 echelle orders, 
spanning from 4580 to 6500~\AA\,(limited by the templates we used), but the complete 
spectrum reaches 8760~\AA.

The CORALIE spectrograph, attached to the 1.2-m Euler telescope in La Silla 
(Chile), works in a simultaneous object-calibration mode, and provides resolution 
of $R\sim70\,000$. Additional ThAr exposures with both fibres are done every 
1-1.5 hours. For this study we used the instrument when it was still
equipped with circular fibres (currently octagonal). Spectra were reduced 
with the dedicated python-based pipeline \citep{jor14,bra17}, which also performs 
barycentric corrections. The pipeline is optimized to derive high-precision radial 
velocities, down to $\sim$5 m/s, and reduces the spectrum to 70 rows spanning 
from 3840 to 6900~\AA. For our purposes, we use only 45 rows (4400--6500~\AA), 
due to the limits of our template spectra and very low signal in the blue part.

Operations at the MPG-2.2m telescope (La Silla, Chile) with the
FEROS instrument \citep{kau99} look very similar to CORALIE, 
as the spectrograph also works in a simultaneous object-calibration 
manner, but employs an image slicer, which gives $R\sim48\,000$, 
and the highest efficiency of all the optical instruments we 
used for this study ($>$20\%), thus provides data with the highest
SNR. Spectra were reduced with the CORALIE 
pipeline adopted to the FEROS data, capable of providing RVs with 
the precision of 5-8 m/s. Although the original spectral format
reaches beyond 10\,000~\AA, the output is reduced to 21 rows covering
4115-6519~\AA, of which we use 20 (4135--6500~\AA).

The three instruments described above were the main source of optical
data, which were later supplemented with observations from two more 
facilities. The observations at Okayama Astrophysical Observatory (OAO)
1.88-m telescope in Okayama (Japan) with the HIDES spectrograph 
\citep{izu99,kam13} were conducted in the fibre mode with image 
slicer ($R\sim50\,000$), without I$_2$, and with ThAr lamp frames
taken every 1-2 hours. The spectra are composed of 
62 rows covering 4080--7538~\AA, of which we use 30 (4365--6440~\AA).
Detailed description of the observing procedure, data reduction and 
calibrations is presented in \citet{hel16}. The
precision reached with our approach is 40-50 m/s.

Finally, several spectra were taken with the HARPS-N spectrograph,
attached to the 3.6-m TNG located at the Roque de Los Muchachos
Observatory in La Palma (Spain). As CORALIE and FEROS, it also works
in a simultaneous object-calibration mode. The data are reduced on-the-fly with the local Data 
Reduction Software, that takes care of the barycentric corrections.
The product is a 1-D spectrum of $R\sim115\,000$ stretching from 
3830 to 6900~\AA. We limit our work to 3850--6500~\AA\,range, 
divided into chunks of 50~\AA\,each.

Below we summarize the optical spectroscopic observations, separately 
for each target.
\begin{itemize}
\item {\it ASAS-052} was observed mainly with CHIRON. Between December
2012 and December 2014 a total of 23 spectra were taken. Another 13 
observations came from CORALIE, which observed ASAS-052 between November 
2013 and November 2014. For none of these we could find any marks of the
secondary of the eclipsing pair with our approach (see Sect.~\ref{sec_rv}), 
so going to the IR was absolutely necessary for this interesting system.
\item {\it ASAS-065} was first observed with FEROS (in March and August
2013), at which four high-SNR spectra were secured. Three more observations, 
done in September and October 2013, come from CHIRON. In December 2013
we started to observe it with HARPS-N, and took six spectra by February 2015.
Seven observations come from CORALIE, and were taken between March 2014 and 
March 2015. Uncertain marks of the secondary were found on about half of the
spectra (most reliable on those coming from FEROS), but their significance
was comparable to artefacts produced by TODCOR. Therefore we decided to 
observe ASAS-065 in IR in order to confirm or reject these detections.
\item {\it ASAS-073} was the first object from this sample we have observed.
Six FEROS spectra were taken between November 2011, and May 2013. 
Later, (January 2015 to January 2017) the star was re-observed with HIDES, 
where six more exposures were taken. On the majority (but not all) of FEROS 
spectra, the secondary was found rather easily, comparing to the two other 
systems, but still somewhat doubtful. Moreover, the HIDES detections were 
marginal, if any. We treated this object as a (successful) test of the 
IR observations, in order to verify if the IR and optical data give consistent 
results.
\end{itemize}

\subsection{Subaru/IRCS infra-red spectroscopy}
As the optical spectra were insufficient for secure identification of
the secondaries, we observed all targets in infra-red. 
We used the 8.2-m Subaru Telescope equipped with
the InfraRed Camera and Spectrograph \citep[IRCS;][]{kob00} working
in the echelle mode, set to the $J$-band. In such setting the instrument
produces spectra composed of 9 echelle orders, stretching from 11616 to 
14285 \AA. We used the narrowest 0.14" slit, which provides the resolution 
of $\sim$20000. Atmospheric turbulence correction was performed with the 
Subaru's AO188 adaptive optics system \citep{hay08,hay10}.
To reduce the influence of the unstable sky background, we observed in 
the ABBA sequence, which is nodding along the slit. Because
it was possible to resolve the two visual components of ASAS-052, we set 
the slit's position angle to $\sim$66.4$^\circ$ so the spectra of both
components were recorded. In further processing we were able to 
extract them separately. 

For the data reduction we used the standard procedures of the 
{\it ccdred} and {\it echelle} packages of the IRAF software.
Flat-fielding was done on the basis of He lamp exposures. First a series 
of frames He-ON was taken and combined, followed by a series of He-OFF. 
Final flat was obtained by subtracting the master OFF from the master ON
frame. Wavelength calibration was based initially on Ar lamp exposures 
taken at the beginning or end of the night, in a similar ON/OFF manner as 
the flats. Later, the velocity corrections were found for each science 
observation separately, from cross-correlation of the science spectra 
with a telluric standard star. Same standards were used to remove the 
strong telluric lines from the spectra. For the cross-correlation itself
we used the task {\it xcsao} of the IRAF's {\it rvsao} package. The
barycentric time and velocity correction were calculated with the {\it bcvcor}
task of the same package. Some RV standards were observed, in order 
to monitor the instrument's stability. Extraction of the ABBA spectra
was done in few steps. First, frames from the position B1 were subtracted
from frames from the position A1 (A1-B1=AB1), and analogously for A2-B2=AB2.
AB1 and AB2 were then added to get the frame AB, which then was also multiplied
by -1 to get the frame BA. The ''positive'' spectra from AB and BA
(corresponding to observations at positions A and B, respectively) were
extracted with the task {\it apal} and wavelength-calibrated with 
{\it refspec} and {\it dispcor}, on the basis of Ar lamp frames.
Location of each ``positive'' spectrum from AB and BA frames served as 
a reference for the Ar lamp extraction. Finally the {\it sarith} task was used
to add AB to BA in the wavelength domain, and obtain the final spectrum.

IRCS observations took place during two nights in March and one night in 
December 2014. ASAS-052 was observed four times, and in three cases the two
visual components were resolved and their spectra extracted separately.
Thanks to this, when the velocities were measured, it was possible to 
determine that the component A is the eclipsing one.
ASAS-065 was observed three times, and ASAS-073 only two times, as the
detections of the secondary were more reliable in the high-SNR FEROS spectra
than for the two other systems. In all cases the cool components 
appeared much more prominent than in the optical data. However, due to a lower
resolution and worse stability, the error bars of the resulting RVs of the 
hotter components are larger than for the optical.

\subsection{Radial velocities}\label{sec_rv}
For the RV measurements we used our own implementation of 
the aforementioned TODCOR technique \citep{zuc94}. 
As templates for the optical data we used synthetic spectra computed 
with ATLAS~9 \citep{kur92}, which do not reach wavelengths 
longer than 6500~\AA. For the IR observations we used synthetic spectra
from the library of \citet{coe05}, which cover both optical and IR regions
(3800-18000\AA), but have lower resolution than the ones from ATLAS~9. 
This is however not a problem for IRCS, which resolution is $\sim$2-6 times 
lower than of other instruments we used.

In this study we had to deal with data with one, two or three sets of 
spectral features, while the TODCOR is optimized for double-lined 
spectroscopic binaries. If the templates are matched correctly, it also
calculates the intensity ratio of the two stars. For single-line cases 
(many of the ASAS-065 and ASAS-073 spectra) we forced the code to look for a 
local maximum around a position where $v_1$ was close to the true velocity 
value, and $v_2$ was very far from it ($>$100~km/s). The code looks for 
the maximum in each axis independently, so the resulting $v_1$ was not affected 
by $v_2$. This approach has been shown to provide satisfactory results in 
\citet{hel16}. 

In the optical spectra, ASAS-052 appeared as either double-
or triple-lined. The maximum of the CCF corresponding to the
Aa component is the strongest one in this system, followed by Ba, and
Bb is the weakest. Initially, the period of the Ba+Bb pair
was not not known, and very often one of its lines was blended with Aa, 
with the other at different velocity, possible to measure. Sometimes they 
were blended with each other, and we could measure the velocity of Aa.
In both such cases, we treated such spectra as double-lined, but in
further analysis we used only the measurement of the unblended component.
When all three sets of lines were seen separately, we run TODCOR twice:
once forcing it to look around the local maximum corresponding to
the positions of Ba and Bb (weaker lines), and the second time for 
the global maximum, corresponding to velocities of Aa and Ba (but
this measurement of Ba was not taken for the orbital fit). Similar
approach was successfully tested on a triple-lined system
\object{KIC~6525196} in \citet{hel17a}, where the $rms$ of the orbital fit was
comparable with the spectrograph's stability measured from standard stars. 

All but one of the IRCS spectra were double-lined, even for ASAS-052. 
As said before, in three of four cases we could extract spectra of 
ASAS-052A and B separately. In the fourth case, the lines of Ba 
overlapped with the lines of Aa, but the features of Ab and Bb were
well separated. We thus ended up with a triple-line case, and measured
the velocities of Ab and Bb only.

Individual RV measurements, their errors, residuals of the fit, as well as 
exposure times, and SNR, are listed in Tables~\ref{tab_rv_052}--\ref{tab_rv_073}
in the Appendix.

\subsection{ASAS photometry}

The All-Sky Automated Survey \citep[ASAS;][]{poj02}
is the first source of photometry for all the targets in our spectroscopic survey.
The data can be extracted from three different sites:
\begin{itemize}
\item{The ASAS Catalogue of Variable Stars 
(ACVS)\footnote{\tt http://www.astrouw.edu.pl/asas/?page=acvs} is the catalogue 
of time-series photometry and identifications of the variables found by ASAS South 
by the third stage of the project (end of 2009). Measurement are made in the 
Johnson's $V$ band. All the studied targets have data obtained from the ACVS.}

\item{ASAS for SuperNovae \citep[ASAS-SN;][]{sha14,koc17} Sky 
Patrol\footnote{\tt https://asas-sn.osu.edu/} is part of the northern counterpart of the 
ASAS project. The $V$-band brightness measurements taken after December 2011 
are available on-line. Many brighter objects suffer from saturation effects, 
and this was the main reason why in this study we used ASAS-SN data only for ASAS-052.}

\item{A number of bright ASAS South variables can be found in the catalogue of 
\citet{sit14}\footnote{\tt http://www.astrouw.edu.pl/$\sim$gp/asas/AsasBrightI.html},
which contains light curves in the $I_{\rm C}$ band. In this archive we only found data for
ASAS-073.}
\end{itemize}

We only used the points that were flagged in the archives as ``good''.
Each archive gives measurements of a certain object from several different cameras 
of a given ASAS station. To correct the systematic differences in the magnitude 
zero points of each camera, we first calculated average values out of the eclipses, 
and later shifted the data accordingly. In cases the number of observations from 
one camera was very low, we excluded them all. For ASAS-052 we merged ACVS and ASAS-SN 
data into one light curve. We also removed obvious outliers. 
In total, we have used 1397, 579 and 550 $V$-band measurements, 
for ASAS-052, ASAS-065 and ASAS-073, respectively, and additional 196 $I_{\rm C}$-band 
points for ASAS-073.

We should also note that some photometric measurements exist in the 
Northern Sky Variability Survey \citep[NSVS;][]{woz04}, but the number 
is low and not sufficient for light curve analysis.

\subsection{MITSuME photometry}\label{sec_mits_photo}
Initial analysis of the public ASAS data showed that the uncertainties of
resulting parameters were very large, mainly because depths of secondary 
eclipses\footnote{We define the primary eclipse as the deeper one, 
and during which the primary component (here also hotter) is eclipsed.} 
are comparable to the scatter of data. Problems with large uncertainties, rising from such a 
situation are presented for example in \citet{cor15} for V1200~Cen. 
Therefore, we decided to obtain additional photometry.

We used the Okayama station of the Multicolor Imaging Telescopes for Survey 
and Monstrous Explosions \citep[MITSuME;][]{kot05} network. This observatory
consists of a 0.5-m robotic telescope, equipped with a multi-band imager 
(Sloan $g'$, Cousins $I_{\rm C}$ and $R_{\rm C}$)
with a field of view of $26'\times26'$. Observations were made between November
2015 and April 2016 in a queue mode, with our targets scheduled in 1-2 hour long 
blocks. Higher priority has been set for the eclipses (especially moments of entry, 
exit, and the minimum), and for some phases out of eclipses. Such scheduling 
coupled with highly unstable weather, unfortunately, turned out to cause problems in 
covering the eclipses of ASAS-052 ($P\simeq5.27$~d) and ASAS-065 
($P\simeq8.22$~d), which last several hours. They are not observable 
every cycle, as they often occur during the daytime. For this reason we do not 
have the minimum and egress of the primary eclipse of ASAS-052, and the ingress 
of the secondary of ASAS-065. Much shorter eclipses of ASAS-073 
($P\simeq1.45$~d) are sampled sufficiently. 

The CCD reduction of the raw data was done with
the standard {\it imred.ccdred} routines under IRAF. Aperture photometry was done
with IRAF's {\it apphot} on the variable and several comparison stars. The object
of interest was usually the brightest star in the field. For the final differential
photometry we chose the comparison star that produced the smallest scatter outside 
of the eclipses, taking into account only frames taken under good weather conditions. 
If more than one comparison gave similar results, the brightest one was chosen.
We have later removed the obvious outliers and measurements with higher formal errors
(normally taken under worse weather conditions, and producing larger scatter).
Due to the shutter failure in the $g'$ camera we had to exclude some observations
in this band. We have also decided not to use $g'$ data for ASAS-052 entirely, 
as this band suffers from the strongest systematics, which rendered the measurements
taken during the secondary eclipse useless (the other two objects have primary 
eclipses well covered, so the use of $g'$ data was still justified). Moreover, 
for ASAS-052 we had no independent information on the amount of third light in $g'$.

The apparent magnitudes of the variable and comparison stars were obtained from the 
available literature sources. In case of ASAS-052 we took $g',r'$, and $i'$ measurements 
from the Sloan Digital Sky Survey (SDSS) data release 12 \citep{ala15}, and 
transformed to the Cousins $R_{\rm C}$ and $I_{\rm C}$ magnitudes using the formulae from
\citet{jor06}. Fields of the other two systems were not observed by the SDSS, so 
we used entries in the same bands from the AAVSO Photometric All Sky Survey (APASS) 
data release 9 \citep{hen15}, and the same transformations to the Cousins' system.
This procedure gave a very good match between MITSuME and ASAS $I_{\rm C}$ light curves
of ASAS-073.

\subsection{Astrometry of ASAS-052~AB}

\begin{table}
\centering
\caption{Results of SOAR speckle observations of ASAS-052.}\label{tab_soar}
\begin{tabular}{ccccc}
\hline \hline
Date &	$\rho$ & $\theta$ & $\Delta mag$ & Band \\
(Besselian years) & [mas] & [$\circ$]	& [mag] & \\
\hline
2015.0288 & 370.3$\pm$0.3 & 67.6$\pm$0.2 & 0.30 & $I$ \\
2015.0288 & 370.4$\pm$0.2 & 67.6$\pm$0.4 & 0.40 & $R$ \\
2015.0288 & 370.4$\pm$0.3 & 67.6$\pm$0.6 & 0.49 & $V$ \\
2017.8269 & 388.8$\pm$0.4 & 65.7$\pm$1.2 & 0.52 & $I$ \\
2017.8269 & 390.3$\pm$0.4 & 65.8$\pm$0.3 & 0.68 & $V$ \\
\hline
\end{tabular}
\tablefoot{ Adopted magnitude differences: $V$: 0.59, $R$: 0.40, $I$: 0.41 ($\pm0.1$~mag for all).}
\end{table}

The WDS currently contains 18 archival astrometric measurements 
of ASAS-052 (WDS~05278-0400), spanning 115 years (1902--2017),
with a large gap between 1978 and 2015.
Orbital motion of the two spectroscopic pairs around their common centre of mass 
can be clearly seen. During the first Subaru run \citep[March 2014, see Figure~1 in][]{hel15} 
we noticed that the secondary is located at a similar position angle as in 1902, which could mean 
that the system has almost completed one revolution since the discovery. 
This was confirmed by subsequent speckle interferometric and lucky imaging observations 
with SOAR taken in 2015 and 2017, which are summarized in Table~\ref{tab_soar}. 
All new data are already published \citep{tok15,tok18},
and are currently included in the WDS archive. Apart from the relative position, also
the magnitude difference between the two components has been measured in three bands.
They were used in the light curve analysis as starting points for the amount of third light. 
The observed discrepancies in $\Delta mag$ between two epochs are normal for speckle data.
For $V$ and $I$ bands we used average $\Delta mag$ values, and set a conservative
uncertainty of 0.1~mag for all bands.


\section{Analysis}

\subsection{Radial velocity curves}

The model RV curves were found with the V2FIT code \citep{kon10}, which fits
a double-Keplerian orbit to a set of RVs of either one or two components of
a spectroscopic binary by $\chi^2$ minimization with a Levenberg-Marquardt
algorithm, and also deals with trends and periodic variations in RVs of a binary.
The code finds the best-fitting two semi-amplitudes $K_{1,2}$, 
systemic velocity of the primary $\gamma_1$, difference in systemic velocities 
of the secondary and primary $\gamma_2-\gamma_1$, eccentricity $e$, argument of 
periastron $\omega$ and the time of phase zero, $T_p$, which is defined here either as
time of periastron passage (for $e>0$) or quadrature (for $e=0$). If at the first run $e$ 
was found not significantly different from zero, then another fit was made with this 
parameter held fixed at 0. The V2FIT can also find the orbital period, but for 
these systems we used values found in JKTEBOP (See next Section). It can also 
fit offsets between measurements from two or more spectrographs, separately for 
each component, as they may vary with the template used for RV measurements.

Due to the fact that we have fewer RV measurements for the secondaries than
for the primaries, and that they come with larger errors, we were analysing
the RVs in two stages. First, each system was treated as an SB1, and only
$K_1, \gamma_1, e, \omega$ and zero-point offsets between spectrographs 
were found from the primary's RVs. Treating the offsets as independent variables 
in the fitting process is necessary to obtain correct and precise values of 
parameters, mainly $K$ and, subsequently, $M\sin^3(i)$, and do not introduce unnecessary 
sources of errors. In the second stage all these parameters were held fixed, 
and only $K_2$ and $\gamma_2-\gamma_1$ were fitted for. We chose this
two-step approach after runing a number of initial fits comparing this manner to
a single fit for all parameters from all RV data simultaneously. We noted, that
in complete SB2 fits the uncertainties of such parameters like $e, T_{\rm p}$ 
and especially $K_1$, are significantly larger than in SB1 fits. This is presumably 
due to fact that they are influenced by RVs of the secondary, which are of worse 
quality than of the primary. In the most extreeme case, the error of $K_1$ in
ASAS-052~A was about 10 times larger in SB2 fit than in SB1, leading to improbably
large uncertainty in mass.

In all V2FIT runs the uncertainties were estimated using the bootstrap approach, 
and additionally, in the second stage ($K_2$ and $\gamma_2-\gamma_1$ search), 
a Monte-Carlo method was used to properly evaluate the influence
of errors of the fixed parameters on the uncertainties of the fitted ones.
In this way we took into account the possible systematics, for example coming from
the template mismatch, spots, or low number of spectra. We also avoid
unrealistically large uncertainties coming from lower-precision data.
In case of the non-eclipsing ASAS-052~B, where both sets of lines were easily visible, 
and RVs of both components are comparably precise, all parameters were 
found simultaneously.

\subsection{Light curves modelling}

Due to different characteristics of the studied objects and data sets used, 
the light curve (LC) analysis was a bit different for each system. Below we
briefly describe the common parts, and follow with explaining the individual
approaches.

We used two well-known and widely-used LC fitting codes. The first, was 
JKTEBOP v28. \citep{sou04a,sou04b}, which is based on the EBOP program \citep{pop81}.
This fast, geometrical code allowed to assess the general characteristic of a 
given system, and good starting values for the second program -- PHOEBE v0.31a
\citep{prs05}, which incorporates the Wilson-Deviney code \citep{wil71}. The advantage of 
PHOEBE is that it works on all light curves simultaneously, and the incomplete LCs of 
ASAS-052 and ASAS-065 were impossible to analyse separately with JKTEBOP. At no point 
we used RV measurements together with LCs. This is because none of the two LC codes 
(the latest version of JKTEBOP works with RVs as well) allows to fit for different
zero-points of various spectrographs.

Initial JKTEBOP runs were made on the ASAS $V$ curves in order to find the orbital 
periods $P$ of systems, to be used in the RV analysis (see previous Section).
Second JKTEBOP approach was done with values of mass ratio $q$, $e$ and $\omega$
found by V2FIT. In general, we fitted the mid-time of the primary eclipse $T_0$
(time of phase zero for JKTEBOP), 
the sum of the fractional radii $r_1+r_2$, their ratio $k$, orbital inclination $i$,
surface brightness ratios $J$, brightness scales (out-of-eclipse magnitudes in each 
filter) $m_{out}$, and fractional amount of the third light $l_3$, if necessary.
The gravity darkening coefficients and bolometric albedos were 
always kept fixed at the values appropriate for stars with convective envelopes 
\citep[$g = 0.32$, $A = 0.5$; ][]{luc67,ruc69}. The logarithmic limb darkening (LD) 
law was used \citep{kin70} with approximate coefficients taken from the tables of 
\citet{vHa96}. For this we assumed temperatures and gravities expected for
main sequence, solar metallicity stars of the masses found from the RVs. 
We checked that, for ASAS data, the uncertainty coming from using slightly 
inaccurate LD coefficients does not significantly influence the final errors of 
LC-based parameters. 

With PHOEBE we analysed all available LCs simultaneously. We fixed the values of 
$P, q, e, \omega$, $\gamma$ (set arbitrarily to 0, but has no influence if RVs are not
used), $a$ (calculated from $a\sin(i)$ and $i$ from initial JKTEBOP runs), 
and primary's effective temperature
$T_\mathrm{eff,1}$ (initially set to a value expected at the main sequence). The starting values 
of modified Kopal potentials $\Omega_{1,2}$ were calculated by PHOEBE from mass ratio 
and fractional radii $r_{1,2}$ obtained with JKTEBOP. Starting values of several 
other parameters (e.g. $i, l_3$) were also set to those found with the JKTEBOP.
In PHOEBE we fitted for: zero-phase time $T_{ph}$ (which for eccentric orbits is 
different from the eclipse mid-time; see the PHOEBE Scientific 
Reference\footnote{\tt http://phoebe-project.org/static/legacy/docs/
phoebe\_science.pdf}), 
$i$, $\Omega_1$, $\Omega_2$, primary
luminosity levels, and $l_3$ in every band, if necessary. We also fitted for the 
secondary's effective temperature $T_\mathrm{eff,2}$, but in order to obtain temperature 
ratio (as the assumed $T_\mathrm{eff,1}$ was treated as uncertain). LD coefficients were 
automatically interpolated by the code from tables of \citet{vHa96}. 
At the last stage we fine-tuned the solutions, applying the results of spectroscopic
or colour indices analysis, i.e. the primary's $T_\mathrm{eff}$ and metallicity of the system 
(when applicable).

\begin{figure*}
\centering
\includegraphics[width=0.9\textwidth]{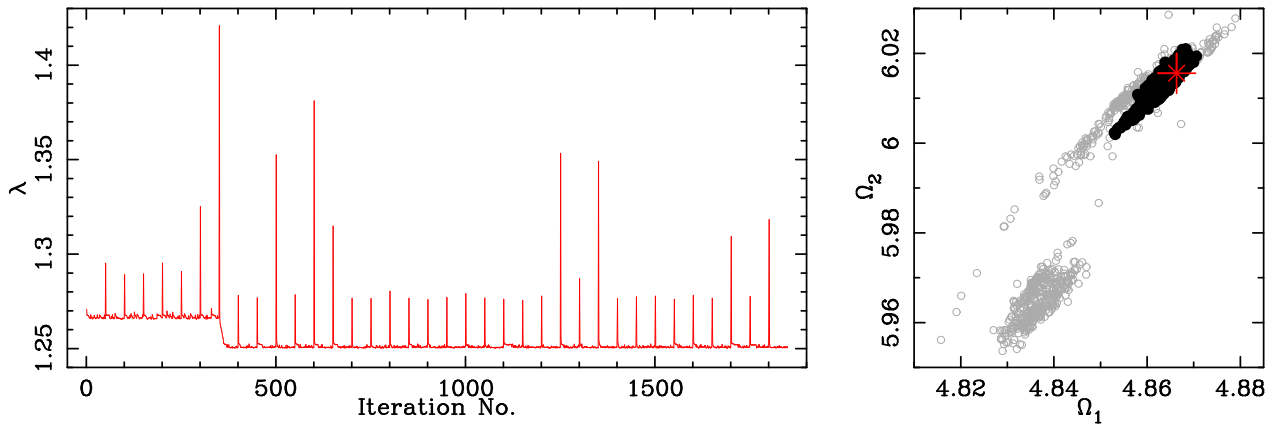}
\caption{Example of the parameter error evaluation with PHOEBE, in the case 
of ASAS-073. {\it Left}: Evolution of the cost function $\lambda$. Peaks represent 
iterations immediately after the kicks. A more significant (deeper, $\lambda\simeq1.25$)
minimum in the parameter hyperspace was found after iteration no. 350. {\it Right:} Mapping the
degeneration between Kopal's modified potentials $\Omega$. The solutions included
in the final error calculation are plotted with black dots, and those that were 
rejected (shallower minimum or still before convergence) are plotted with grey circles.
The red $\times$ symbol with error bars represents the best fit (lowest $\lambda$) 
and the final 1$\sigma$ uncertainties.}\label{fig_err_calc}
\end{figure*}

Parameter errors given directly by PHOEBE have the tendency to be underestimated.
Therefore, to reliably estimate the influence of systematics, and correlations
between different parameters in the hyperspace, we followed the procedure of heuristic
scanning suggested in \citet{prs05}. We run a number ($\sim$2000) of iterations,
recording the output after each one, and introducing ``kicks'' every 50 steps, 
i.e. changing the current values of parameters by 10 times their formal errors in a
random direction. After such kick, the fit converged to a minimum (typically
after 3-5 steps), which was later
evaluated using the value of the cost function $\lambda$ \citep[see: ][]{prs05},
calculated directly by PHOEBE. If several minima were found, we choose the one
with the lowest $\lambda$. To evaluate the final uncertainties, we took into account
results of only those iterations, which ended up near the chosen minimum. For each
parameter, we calculated the $rms$, and added it in quadrature to the formal
fit error. Parameters strongly affected by systematics and degenerations typically 
have these $rms$-es larger than formal fit errors. An example of the procedure is shown
in Figure~\ref{fig_err_calc}.

\subsubsection{ASAS-052}
This system was the most difficult one to model, because of a strong third light, and
poor coverage of the primary eclipse in MITSuME data. For reasons explained earlier,
we did not use MITSuME's $g'$-band observations. Initially, we kept $l_3$ fixed to 
values found from SOAR observations, both in JKTEBOP and PHOEBE. The JKTEBOP fits 
on ASAS $V$-band curve favoured solutions with a total secondary eclipse. In such 
situation, and having $l_3$ estimated, we could independently calculate fractional
fluxes of each eclipsing component. However, while these were incorporated in
PHOEBE, we could not find a satisfactory fit to all LCs simultaneously. We solved
this problem after setting $l_3$ free for the ASAS $V$-band under JKTEBOP.
The solution we found was still showing a total secondary minimum, and predicted 
a slightly larger $l_3$ in $V$ than the SOAR observations. This 
can be explained with the fact that the ASAS photometric aperture and pixel scale
is larger than MITSuME's, and additional background bodies were included. With the
corrected $l_{3,V}$, in PHOEBE we found a satisfactory fit to all the data.
Without independent information about temperatures, in later stages of PHOEBE fit
we decided to decouple the secondary's luminosities from temperatures. This in 
practice means fitting for the fluxes of primary and secondary independently.
This way we found the colour indices for the primary 
and used them to estimate its $T_\mathrm{eff}$ (see further Sections).
 Very large uncertainties of the colours 
of the secondary did not allow for a similar estimate for this component. Instead,
we used the temperature ratio found earlier, and the colour-based $T_\mathrm{eff,1}$.
In the final step we fixed both values of $T_\mathrm{eff}$ and fine-tuned the model.

\subsubsection{ASAS-065}
We initially thought that the eclipses are partial, but MITSuME observations
showed that the shape of the well-covered primary minimum is better explained
with inclination very close to 90$^\circ$, when the secondary star transits
close to the centre of the primary's disk. This configuration also predicted a 
total secondary eclipse, but the depths of both minima could not be explained.
We thus decided to look for the third light, and found a satisfactory solution
with small, but measurable (several per cent) values of $l_3$ in each band.
We were fortunate to find that a small portion of MITSuME observations in the 
secondary eclipse was done during the total phase, which turned out to
be crucial in constraining the ratios of fluxes and radii securely.

\subsubsection{ASAS-073}
This system was relatively easy to model. We used the largest number of light 
curves (five), all with very good coverage in eclipses, and sufficient out of them.
No third light contribution had to be taken into account. This system also shows
a total secondary eclipse (which was obvious from ASAS $V$ data alone).


\subsection{Atmospheric parameters and abundances from spectra}
In ASAS-065 and -073, the primary star dominates the flux in $V$, so it 
is possible to independently assess a number of atmospheric parameters 
(e.g. $T_{\rm eff}, \log(g)$) of the primary, and abundances of elements 
from optical spectra. The same cannot be done for ASAS-052 without 
disentangling the spectrum of Aa from the two other visible components Ba and Bb. 

For the following, we used continuum-normalized FEROS spectra, 
shifted by the value of measured RV of the primary, stacked together, 
and corrected for the additional flux. Influence of additional bodies was assumed 
to be constant.

The effective temperature $T_{\rm eff}$, surface gravity $\log(g)$ and 
microturbulence $\xi_{\rm t}$ of the analysed stars were determined using the 
Balmer lines, neutral and ionised iron lines and strong lines of Mg, Ca and Na. 
For both stars we used spectrum synthesis method relying on an efficient 
least-squares optimisation algorithm \citep[see eg.][and references therein]{nie15}. 
This method allows for a simultaneous determination of various parameters that 
affect the shape of spectral lines, like $T_{\rm eff}$, $\log(g)$, $\xi_{\rm t}$, 
$v\sin(i)$, and the relative abundances of the elements. Because the 
atmospheric parameters are correlated, $T_{\rm eff}$, $\log(g)$, and $\xi_{\rm t}$ 
were obtained prior to the chemical abundance analysis. The $v\sin(i)$ values 
were determined by comparing the shapes of observed metal line profiles with 
the computed profiles \citep{gra05}. 

In later steps, after $T_{\rm eff}$ of the secondaries were estimated, 
the influence of additional components on the atmospheric parameters of primaries 
was verified in a following way. First, a synthetic, continuum-normalized, ``clean'' 
spectrum of a primary was created with SNR representative for the shift-and-stacked 
FEROS spectra (with relative fluxes preserved). Then, a series of synthetic spectra of 
secondaries were created, with relative RV shifts, fluxes and SNRs corresponding
to the FEROS observations. They were summed, scaled by $l_2/l_1$, and added to 
the ``clean'' spectrum of the primary to create ``dirty'' spectra (which mimick the
true shift-and-stacked FEROS spectra used for the analysis). The atmospheric parameter
search has been repeated on both ``clean'' and ``dirty'' spectra (for both 
ASAS-065 and ASAS-073), and results compared. We found out that the ``clean'' 
spectrum gives results indifferent from the ``dirty'' one, i.e. differences are
smaller than half the uncertainties we have adopted.
As for individual lines, the depth changed by no more than 4\%, and, due to 
the varying RV difference between primaries and secondaries, the influence 
on abundances should be negligible. Only the strongest lines of the secondaries
(which are smeared in RV in any case) could affect single lines of the primaries, 
but this would not have changed the overall, final results.

Chemical abundances thenselves were determined 
from many different spectral parts. Every investigated part of the 
spectrum consisted of one line or many lines and blends, depending on the 
analysed wavelength range and the rotation velocity of the star. In the case 
of slowly rotating stars line profile fitting is possible, however for stars 
with medium and high $v\sin(i)$ values, spectrum synthesis on broader spectral 
parts is necessary. At the end of the chemical abundances analysis we derived 
their average values.

We used atmospheric models (plane-parallel, hydrostatic and radiative equilibrium) 
computed with the ATLAS~9 code, and synthetic spectra calculated with the line-blanketed, 
local thermodynamical equilibrium code SYNTHE \citep{kur05,sbo05}. We used the line 
list available at the web-site of F.~Castelli\footnote{\tt http://wwwuser.oats.inaf.it/castelli/}.

\begin{table}
\centering
\caption{Chemical abundances of ASAS-065 and ASAS-073  and their standard deviations, calculated for elements calculated from more than three spectral parts. 
Solar values are taken from \citet{asp09}.}
\label{tab_spanal}
\begin{tabular}{llll}
\hline
Element	&  \multicolumn{3}{c}{Abundances}\\
(Z)	&    ASAS-065  &   ASAS-073  &  Solar \\
\hline
  C   (6)  &   8.68$\pm$0.18(4)          &    8.57$\pm$0.22(5)         & 8.43\\
  Na (11)  &   6.57$\pm$0.08(6)          &    6.15$\pm$0.09(4)         & 6.24\\
  Mg (12)  &   7.71$\pm$0.12(4)          &    7.35$\pm$0.09(6)         & 7.60\\
  Si (14)  &   7.62$\pm$0.26(40)         &    7.26$\pm$0.22(16)        & 7.51\\
  S  (16)  &                             &    7.27(2)                  & 7.12\\
  Ca (20)  &   6.57$\pm$0.19(31)         &    6.26$\pm$0.10(14)        & 6.34\\
  Sc (21)  &   3.37$\pm$0.15(25)         &    2.94$\pm$0.06(7)         & 3.15\\
  Ti (22)  &   5.14$\pm$0.13(148)        &    4.77$\pm$0.17(30)        & 4.95\\
  V  (23)  &   4.19$\pm$0.16(75)         &    4.01(2)                  & 3.93\\
  Cr (24)  &   5.79$\pm$0.17(147)        &    5.51$\pm$0.16(25)        & 5.64\\
  Mn (25)  &   5.55$\pm$0.13(50)         &    5.22$\pm$0.17(9)         & 5.43\\
  Fe (26)  &   7.59$\pm$0.13(387)        &    7.23$\pm$0.08(80)        & 7.50\\
  Co (27)  &   5.01$\pm$0.17(90)         &    5.09$\pm$0.20(6)         & 4.99\\
  Ni (28)  &   6.40$\pm$0.17(133)        &    6.00$\pm$0.17(33)        & 6.22\\
  Cu (29)  &   4.38$\pm$0.21(5)          &    4.08(1)                  & 4.19\\
  Zn (30)  &   4.73(2)                   &    4.18(1)                  & 4.56\\
  Sr (38)  &   2.83(2)                   &    2.85(1)                  & 2.87\\
  Y  (39)  &   2.24$\pm$0.19(12)         &    2.09$\pm$0.22(7)         & 2.21\\
  Zr (40)  &   2.71$\pm$0.33(10)         &    3.12(1)                  & 2.58\\
  Mo (42)  &   2.15(2)                   &                             & 1.88\\
  Ru (44)  &   1.98(1)                   &                             & 1.75\\
  Ba (56)  &   2.42$\pm$0.04(3)          &    2.39$\pm$0.05(3)         & 2.18\\
  La (57)  &   1.27$\pm$0.24(7)          &    1.80(2)                  & 1.10\\
  Ce (58)  &   1.80$\pm$0.27(10)         &    2.29(2)                  & 1.58\\
  Pr (59)  &   0.46$\pm$0.11(3)          &                             & 0.72\\
  Nd (60)  &   1.56$\pm$0.24(19)         &    1.59$\pm$0.07(3)         & 1.42\\
  Sm (62)  &   1.26$\pm$0.32(7)          &	                           & 0.96\\
  Eu (63)  &   1.74(1)                   &                             & 0.52\\
  Tb (65)  &   0.29(1)                   &                             & 0.30\\
  Dy (66)  &   0.79(1)                   &                             & 1.10\\
\hline
\end{tabular}
\\ Note: Numbers in parenthesis in columns 2 and 3 show the number
of individual lines that were used. For less than three lines the 
uncertainties were not estimated.
\end{table}

\begin{figure}
\centering
\includegraphics[width=\columnwidth]{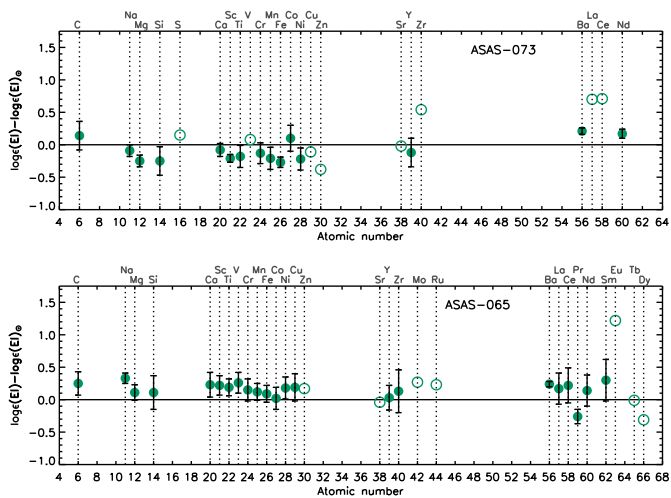}
\caption{Abundances of ASAS-073 (top) and ASAS-065 (bottom)
from the detailed spectral analysis, compared to solar abundances. Open circles without
errorbars refer to elements for which fewer than 3 lines were used, therefore errors
were not estimated.
\label{fig_spanal}}
\end{figure}

For ASAS-065 we used Balmer H$\beta$ and H$\alpha$ lines to find the first 
approximation of effective temperature. In this step $\log(g) = 4.0$\,dex, 
$\xi_{\rm t} = 1$\,km/s and solar metallicity were assumed as starting points. 
For $T_{\rm eff} < 8000$\,K Balmer lines are not sensitive to the $\log(g)$. 
Next, the atmospheric parameters were improved through the analysis of 
\ion{Fe}{i} and \ion{Fe}{ii} lines. Effective temperature was changed until 
there was no trend in the abundance versus excitation potential for the 
\ion{Fe}{i} lines. Microturbulence was adjusted until there was no correlation 
between iron abundances and line depths for the \ion{Fe}{i} lines. 
Surface gravity was found by requiring the same abundances from the analysis 
of \ion{Fe}{i} and \ion{Fe}{ii} lines. Simultaneously, the projected rotational 
velocity $v\sin(i)$ was obtained. Strong lines of \ion{Na}{i} D1 (5889.95\,{\AA}) 
and D2 (5895.92\,{\AA}), \ion{Ca}{i} (6162.18\,{\AA}), and \ion{Mg}{i}\,b 
(5183.62\,{\AA}) show strong pressure-broadened wings in the spectra of 
cool stars, and were used for the determination of surface gravity \citep{gra05}. 
The uncertainty of $\log (g)$ was obtained by changing the previously found 
$T_{\rm eff}$ and $\xi_{\rm t}$ in their error bars. Final atmospheric parameters, 
$T_{\rm eff} = 5500 \pm 100$\,K, $\log (g) = 4.4 \pm 0.1$, $\xi_{\rm t} = 1.0 \pm 0.1$\,km/s, 
and $v\sin(i) = 5.8 \pm 0.4$\,km/s were used to perform the detailed 
analysis of chemical abundances, which is summarized in Table~\ref{tab_spanal} 
and Figure~\ref{fig_spanal}. ASAS-065 appears to be slightly more metal rich
than the Sun.

The initial effective temperature of ASAS-073 was also obtained from 
Balmer H$\beta$ and H$\alpha$ lines, and subsequently improved by the analysis 
of \ion{Fe}{i} abundances dependency with excitation potential. Also the 
microturbulence and the $v\sin{i}$  were determined in the same way as for 
ASAS-063. The determined atmospheric parameters, $T_{\rm eff} = 6400 \pm 100$\,K, 
$\log (g) = 4.2 \pm 0.1$, $\xi_{\rm t} = 1.6 \pm 0.2$\,km/s and $v\sin(i) = 54 \pm 2$\,km/s 
were used to perform the detailed analysis of chemical abundances 
(see Tab.~\ref{tab_spanal} and Fig.~\ref{fig_spanal}). ASAS-073 appears to be 
rather metal-poor in comparison to the Sun.

The determined microturbulence velocities for both stars are typical for 
stars in the observed temperature and surface gravity ranges 
\citep[see eg.][]{sma04,nie15,nie17}. 

The obtained parameters are subject to errors resulting from a number 
of different sources, including those coming from adopted atmospheric 
models calculated with specific assumptions \citep[e.g. LTE instead of full 
non-LTE and 1-D rather than 3-D approach, see ][for more discussion]{nie15}. 
The influence of a selected set of atomic data and their completness have 
to be noted. The important factors are quality and wavelength range of the 
analysed spectrum and its normalisation. The last parameter is particularly 
important for heavily blended spectra of stars with moderate or high 
$v\sin(i)$ values. In addition, the chemical abundance values are 
influenced by inaccurate atmospheric parameters $T_{\rm eff}$, $\log(g)$, 
and $\xi_{\rm t}$. Such errors were discussed by \citet{nie15}. Usually, the 
combined errors of chemical abundances calculated assuming 
$\Delta T_{\rm eff} = 100$\,K, $\Delta \log(g) = 0.1$, 
and $\Delta \xi_{\rm t} = 0.1$\,km\,s$^{-1}$ is less than 0.2\,dex.

In case of ASAS-065 we have also independently estimated $T_{\rm eff}$ using line
depth ratios (LDR). It was possible thanks to low rotational velocity (narrow, 
unblended lines) and absence of well-visible lines coming from the secondary or 
the third light. The same could not be done for ASAS-073 due to significant 
rotational broadening ($>$50~km/s), which made the lines of interest blended,
and the weaker ones not recognizable. We chose five LDR-$T_{\rm eff}$ calibrations
from \cite{kov04}, with the smallest $rms$ (below 50~K), and used them for four 
FEROS and five HARPS-N spectra. The results were averaged, and the $rms$ was 
adopted as the uncertainty. We obtained $T_\mathrm{eff,1}=5420(60)$~K, confirming 
the value obtained from Balmer and Fe lines (which was ultimately taken as the final). 
This agreement again shows that, in our spectral analysis, the contribution of the second 
(and third) light has been sufficiently accounted for, and results are reliable. Further 
confirmation comes from RV+LC estimates of $\log(g_1)$ and velocity of synchronous 
rotation $v_{\rm syn,1}$ (see further Sections), which can be compared to $v\sin(i)$. 
For both ASAS-065 and ASAS-073 the agreement in both parameters is better than 2$\sigma$.

\subsection{Absolute parameters}

The absolute values of parameters were calculated with the 
JKTABSDIM\footnote{\tt http://www.astro.keele.ac.uk/jkt/codes.html}
procedure, which is available with JKTEBOP. This code combines the output of
spectroscopic and light curve solutions to derive a set of stellar absolute dimensions, 
related quantities, and distance, if effective temperatures are known or can 
be estimated. For distance determination, this code uses the apparent, total 
magnitudes of a given binary in $U,B,V,R,I,J,H,K$ bands.  
For ASAS-052 and ASAS-063, we could only use estimates of apparent total
brightness in $V,R$, and $I$, corrected for the influence of the third light. 
In case of ASAS-073, we also used $J,H,K$ from 2MASS \citep{scr06}.
The code compares the observed (total) magnitudes with absolute ones, calculated using
a number of bolometric corrections \citep{bes98,cod76,flo96,gir02}, and
surface brightness-$T_\mathrm{eff}$ relations from \citet{ker04}. 
Flux ratios may also be used to further constrain individual absolute magnitudes of 
each component. As the final distance value, we adopted a weighted average of all 
individual results given by JKTABSDIM.

Apart from stellar, photometric, and orbital parameters, JKTABSDIM also 
calculates the rotation velocities predicted for the case of synchronisation 
of rotation with orbital period $v_{syn}$, the time scale of such synchronisation 
$\tau_{syn}$, and the time scale of circularisation of the orbit $\tau_{cir}$. 
The two time scales are calculated using the formalism of \citet{zah75,zah77}.

\subsection{Comparison with isochrones}

In order to estimate the age and evolutionary status of the studied systems, we
compared our results with theoretical isochrones. We made the comparison on
mass $\mathcal{M}$ vs. radius $R$, effective temperature $T_{\rm eff}$, 
and luminosity $L$ planes. We used the latest version (v1.2S) of PAdova and TRieste 
Stellar Evolution Code\footnote{\tt http://stev.oapd.inaf.it/cgi-bin/cmd} 
\citep[PARSEC;][]{bre12,mar17}. The solar-scaled
composition for this set follows the relation $Y=0.2485+1.78 Z$, and the 
solar metal content is $Z_\odot=0.0152$ \citep{bre12}. Whenever possible, we used our
own estimates of [M/H], assuming it is equal to [Fe/H] from the spectral analysis.


\section{Results}

\begin{figure}
\centering
\includegraphics[width=0.95\columnwidth]{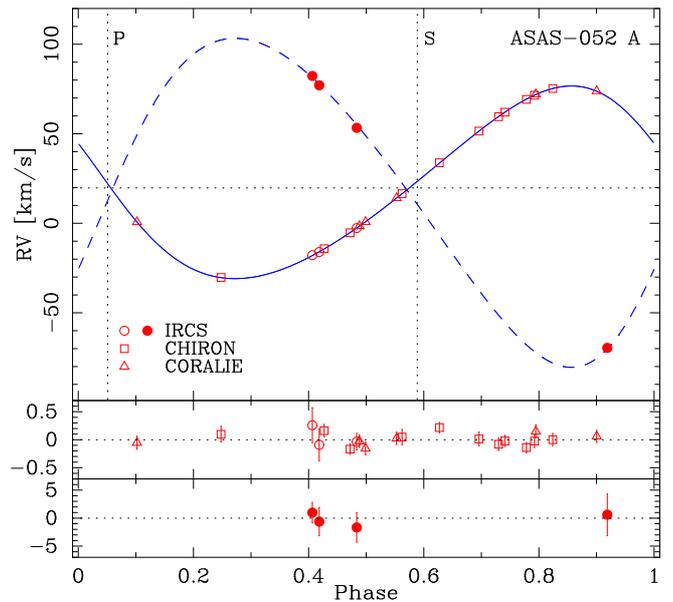}
\caption{Model RV curves (blue lines) and measurements (red points)
of the eclipsing pair ASAS-052~A. Open symbols and the solid line refer to the primary,
and solid symbols and the dashed line to the secondary. The dotted horizontal line marks the
systemic velocity of the primary. Residuals (all in km/s) are shown below.
Symbols used for each of the instruments are labelled. The
phase zero is set to the pericentre time $T_P$. Phases of 
primary (P) and secondary (S) eclipses are marked with vertical dotted lines 
and labelled.\label{fig_rv_A052A}}
\end{figure}

\subsection{ASAS-052}

This is a rare example of a high-order ($n>3$) multiple that contains an eclipsing binary.
There are only several such systems known, including YY~Gem, \citep{tor02},
\object{V994~Her} \citep{lee08}, \object{ASAS~J011328-3821.1} \citep{hel12}, 
\object{HD~86222} \citep{dim14}, \object{KIC~7177553} \citep{leh16}, 
1SWASP~J093010.78+533859.5 \citep{koo14}, \object{EPIC~220204960}~\citep{rap17}, 
\object{V482~Per}~\citep{tor17}, \object{KIC~4150611} \citep{hel17b}, or 
EPIC~219217635 \citep{bor18}. Such systems are usually found in hierarchical 
configurations, meaning that their components tend to form short-period pairs, which 
themselves are in large separations. Such objects are interesting from the point of view 
of stellar formation and dynamical interactions. 

Below we separately discuss our results for the eclipsing and non-eclipsing pairs,  
and for the ``large'', astrometric orbit.

\begin{table*}
\centering
\caption{Orbital, photometric, and absolute physical parameters of the three studied eclipsing binaries}\label{tab_par_all}
\begin{tabular}{lccc}
\hline \hline
Parameter & {\bf ASAS-052~A} & {\bf ASAS-065} & {\bf ASAS-073} \\
\hline
$P$ [d]				& 5.2735660(9)	& 8.219626(13)	& 1.446253(2)	\\
$T_0$ [JD-2450000]	& 1873.2045(7)	& 1883.320(2)	& 1870.166(2)	\\
$T_{\rm p}$ [JD-2450000]	& 2874.920(4)	& 1887.92(6)	& 1869.803(2)	\\
$K_1$ [km/s]		&  53.8(3)		&  51.32(4)	&  88.3(5)	\\
$K_2$ [km/s]		&  91.9(2.3)	&  72.84(4)	& 158.6(1.6)	\\
$q$			& 0.586(15)	& 0.705(4)	& 0.556(6)	\\
$e$			& 0.145(5)	& 0.0113(7)	& 0.0(fix) 	\\
$\omega$ [$^\circ$]	&  66(2)	& 291(3)	& --- \\
$\gamma_1$ [km/s]	& 19.8(7)	& 10.36(4)	& 35.1(4)	\\
$\gamma_2-\gamma_1$ [km/s] & $-$3(7)	& 0.3(4)	& 2.8(1.4)	\\
$\mathcal{M}_1 \sin^3{i}$ [M$_\odot$]	& 1.03(6)	& 0.956(12) & 1.448(34) \\
$\mathcal{M}_2 \sin^3{i}$ [M$_\odot$]	& 0.60(2)	& 0.674(5)  & 0.806(13) \\
$a \sin{i}$ [R$_\odot$]			& 15.03(24)	& 20.18(7)  & 7.06(5)	\\
$rms_{\rm RV,1}$ [km/s]\tablefootmark{a}	& 0.105/0.114	& 0.019/0.100	& 1.00/1.35	\\
$rms_{\rm RV,2}$ [km/s]\tablefootmark{a}	&   ---/1.24	& 0.255/1.168	& 4.09/4.54	\\
$i$ [$^\circ$] 	& 88.94(5)  	& 89.66(2) 	& 87.91(1)	\\
$r_1$			& 0.0682(6) 	& 0.0494(1)	& 0.2314(6)	\\
$r_2$			& 0.0390(13)	& 0.0342(3)	& 0.1160(14)	\\
$\Omega_1$	& 15.24(12)	& 20.93(4)	& 4.866(4)	\\
$\Omega_2$	& 16.22(13)	& 21.73(7)	& 6.016(5)	\\
$T_{\rm eff,2}/T_{\rm eff,1}$\tablefootmark{b} & 0.697(2)	& 0.722(1)	& 0.6835(6) \\
$(l_2/l_1)_{\rm ASAS\;V}$	& 0.018(12)	& 0.06(5)	& 0.032(6)    \\
------$_{\rm ASAS\;I}$	& --- 			& --- 		& 0.060(14)   \\
------$_{\rm MITSuME\;g'}$	& --- 		& 0.032(5)	& 0.0294(5) \\
------$_{\rm MITSuME\;R}$	& 0.020(2)	& 0.086(5)	& 0.0467(2) \\
------$_{\rm MITSuME\;I}$	& 0.054(4)	& 0.136(5)	& 0.0604(3) \\
$(l_3)_{\rm ASAS\;V}$	& 0.44(3)	& 0.09(4)	& --- \\
---$_{\rm MITSuME\;g'}$	& --- 		& 0.084(4)	& --- \\
---$_{\rm MITSuME\;R}$	& 0.41(2)	& 0.092(4)	& --- \\
---$_{\rm MITSuME\;I}$	& 0.41(2)	& 0.010(4)	& --- \\
$rms_{\rm ASAS\;V}$ [mag] 		& 0.021	& 0.018	& 0.015	\\
------$_{\rm ASAS\;I}$ [mag]	& --- 	& --- 	& 0.021	\\
------$_{\rm MITSuME\;g'}$ [mag]	& --- 	& 0.019	& 0.016	\\
------$_{\rm MITSuME\;R}$ [mag] 	& 0.015	& 0.016	& 0.013	\\
------$_{\rm MITSuME\;I}$ [mag] 	& 0.017	& 0.016	& 0.015	\\
$\mathcal{M}_1$ [M$_\odot$] & 1.03(6)	& 0.956(12)	& 1.452(34)	\\	
$\mathcal{M}_2$ [M$_\odot$] & 0.60(2)	& 0.674(5)	& 0.808(13)	\\	
$a$ [R$_\odot$]   	& 15.03(24)	& 20.18(7)	& 7.06(5)	\\
$R_1$ [R$_\odot$] 	& 1.03(2)	& 0.997(4)	& 1.635(12)	\\	
$R_2$ [R$_\odot$] 	& 0.59(2)	& 0.690(7)	& 0.819(11)	\\	
$\log(g_1)$       	& 4.430(14)	& 4.421(3)	& 4.173(5) 	\\
$\log(g_2)$       	& 4.684(28)	& 4.589(9) 	& 4.519(10)	\\
$v_{\rm syn,1}$ [km/s] 	&  9.8(2)	& 6.14(3)	& 57.2(4)	\\
$v_{\rm syn,2}$ [km/s] 	&  5.6(2)	& 4.25(5)	& 28.7(4)	\\
$\log(\tau_{\rm syn})$ [yr]	& 7.152(14)	& 7.825(3)	& 4.932(6)	\\
$\log(\tau_{\rm cir})$ [yr]	& 9.194(8)	&10.276(2)	& 6.183(3)	\\
$T_{\rm eff,1}$ [K]\tablefootmark{c}	& 5300(340)	& 5500(100)	& 6400(100)	\\
$T_{\rm eff,2}$ [K]\tablefootmark{c}	& 3700(230)	& 3970(110)	& 4370(110)	\\
$\log(L_1/L_\odot)$	& $-$0.13(11)	& $-$0.09(3)	&    0.60(3)	\\
$\log(L_2/L_\odot)$	& $-$1.24(11)	& $-$0.96(5)	& $-$0.66(5)	\\
$[Fe/H]$		& +0.5\tablefootmark{d}	& +0.09(13)	& $-$0.27(8)	\\
$d$ [pc]\tablefootmark{e}	& 102(14)	& 108(10)	& 170(5)	\\
\hline
\end{tabular}
\tablefoot{
\tablefoottext{a}{Without and with IRCS.}
\tablefoottext{b}{Obtained with PHOEBE.}
\tablefoottext{c}{From various sources, see text for details.}
\tablefoottext{d}{Assumed.}
\tablefoottext{e}{Without extinction}
}
\end{table*}

\begin{figure*}
\centering
\includegraphics[width=0.7\textwidth]{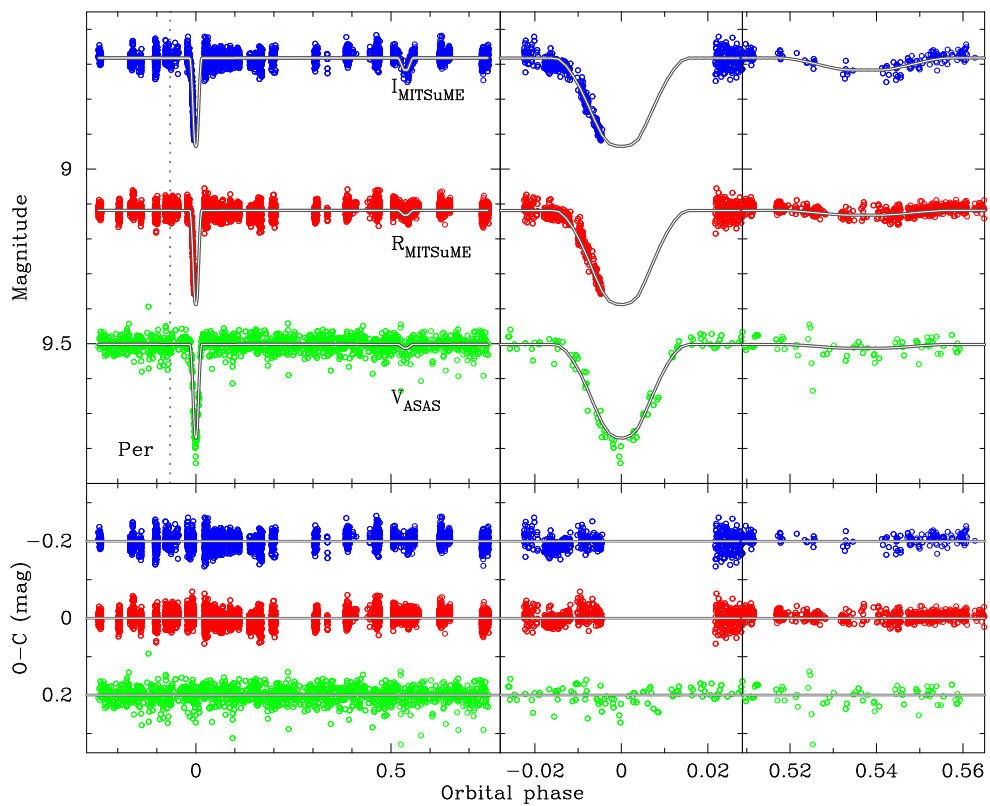}
\caption{Model light curves (white lines) and measurements (coloured points)
of ASAS-052. The Y axis shows the observed magnitude, with no 
vertical shifts. Bands and instruments are labelled. Residuals (O-C) are shown below.
For consistency, phase zero is set to the primary eclipse mid-time $T_0$. 
The phase of pericentre passage is marked with vertical dotted line and labelled~``Per''. 
Note how the depth of the secondary eclipse changes with wavelength.\label{fig_lc_A052}}
\end{figure*}

\subsubsection{The eclipsing pair}

In Figure~\ref{fig_rv_A052A} we present the observed and model RVs of 
ASAS-052~A. The MITSuME $I_{\rm C},R_{\rm C}$, and ASAS $V$-band photometry, with the 
best-fitting model light curves, are shown in Figure~\ref{fig_lc_A052}. 
Orbital and physical parameters are shown in Table~\ref{tab_par_all}.
As mentioned before, the secondary was seen only in the IR spectra. 
Note that this pair has the lowest flux ratio $l_2/l_1$ in all bands 
(0.017 in $V$), and the secondary is the lowest-mass star in our sample 
(0.604~M$_\odot$). Low number of its RV measurements (four) is the main 
reason of large mass errors -- 5.72 and 3.31 \% for the primary and 
secondary respectively. It is still reasonably good, but far from the state 
of the art. Relative errors in radii are 1.85 and 3.74\% for the primary 
and secondary, respectively, and are satisfactory, considering incomplete minima 
coverage. The errors in fractional radii come from the uncertainty of Kopal's potentials 
$\Omega$, which themselves were degenerated with inclination, and the mass
ratio, which is embedded in $\Omega_{1,2}$. These
degenerations are taken into account in the adopted errors.

The primary absolutely dominates the pair A, and constitutes about 55\% of the total flux of 
the whole quadruple system, while over 40\% comes from the additional light. For this reason 
we did not take into account any estimates of the effective temperature found in the literature, 
which all assume the target is a single star. As mentioned before, we have calculated three 
observed colours, and used the colour-$T_\mathrm{eff}$ calibrations from \citet{wor11}. 
We adopt a weighted average of single $T_\mathrm{eff}$ values predicted by each colour 
index. 
For the primary we got $(V-I_{\rm C})_1 = 0.83(8)$, $(V-R_{\rm C})_1=0.43(8)$, 
and $(R_{\rm C}-I_{\rm C})_1 = 0.41(6)$~mag, which together gave 5300(340)~K, assuming $\log(g)_1=4.408$~dex.
Analogously, for the secondary we have $(V-I_{\rm C})_2 = 2.1(9)$, $(V-R_{\rm C})_2=0.5(9)$, 
and $(R_{\rm C}-I_{\rm C})_1 = 1.50(15)$~mag. The errors of $(V-I_{\rm C})$ and $(V-R_{\rm C})$ are so large because 
the uncertainty of the secondary's flux in $V$ is comparable to the flux
itself -- 0.20(13), in PHOEBE's arbitrary units. This error is basically the same as for 
the flux of the primary -- 11.02(13). In cases of pairs with such a high contrast, 
the poor quality of ASAS photometry can even make the error larger than the 
contribution of the fainter star.

\begin{figure}
\centering
\includegraphics[width=0.8\columnwidth]{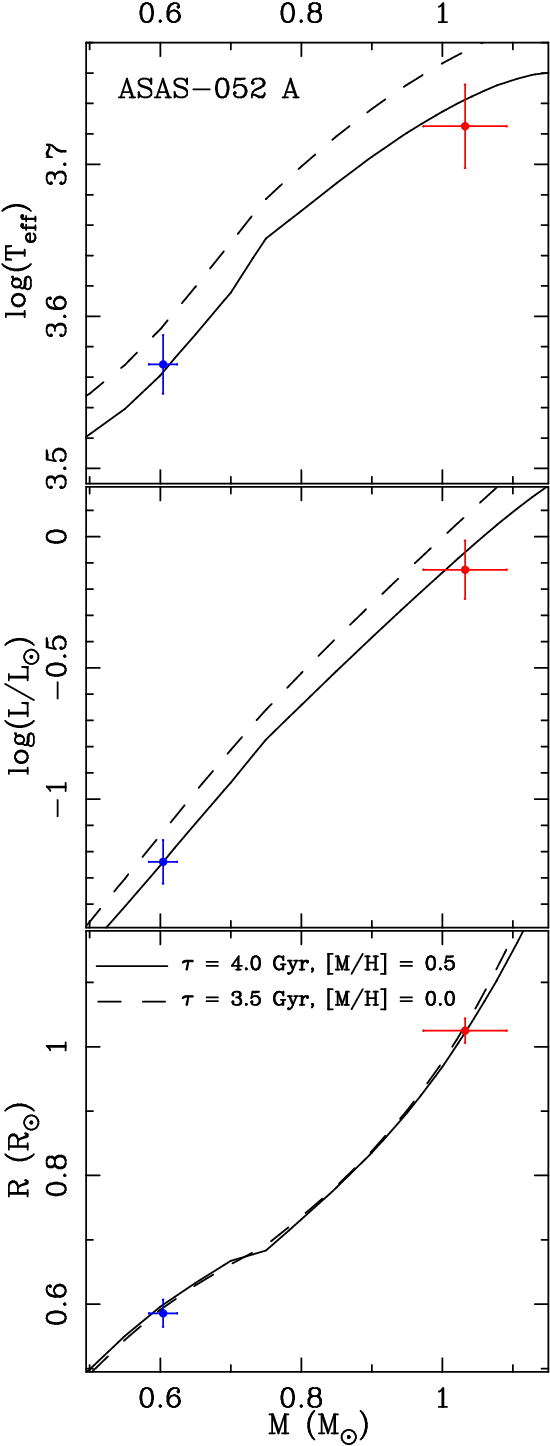} 
\caption{Comparison of our results for ASAS-052~A with theoretical PARSEC
isochrone for the age of 4.0~Gyr and [M/H]=+0.5~dex (solid line) and
solar-composition 3.5~Gyr (dashed)} on mass vs. temperature (top),
luminosity (middle), and radius (bottom) planes. The primary is marked with the red
symbol, and the secondary with blue. \label{fig_iso_A052}
\end{figure}

For the record, $T_{\rm eff,2}$ estimated from $(R_{\rm C}-I_{\rm C})$, the only index which 
had formal uncertainties sufficiently small, is 3180(120)~K. The primary's temperature
from $(R_{\rm C}-I_{\rm C})$ only is 5170(380)~K. Ratio of these two colour-based temperatures 
 $T_{\rm eff,2}/T_{\rm eff,1} = 0.62\pm0.05$ -- agrees within 2$\sigma$ with the one
obtained from PHOEBE -- $0.697\pm0.002$.
Both values are likely affected by the incomplete coverage 
of primary eclipse in MITSuME data. 

The distance obtained with JKTABSDIM, based on the adopted temperatures 
(Tab.~\ref{tab_par_all}), is 102(14)~pc. No interstellar extinction has 
been assumed. We used the following observed magnitudes:
10.12(7)~mag in $V$, 9.69(4)~mag in $R_{\rm C}$, and 9.25(5)~mag in $I_{\rm C}$.
The resulted value is the weighted average of nine single estimates, and their $rms$
is taken as the distance uncertainty. Unfortunately, this system does not have any 
direct parallax determination, even from GDR2, probably because it is a visual binary,  
so we were not able to confirm if our temperature scale is sound.

The temperatures were also crucial for the comparison with isochrones, which is shown in 
Figure~\ref{fig_iso_A052}. Due to large uncertainties of $T_{\rm eff}$, and lack of 
other, independent estimates of temperatures, the following results should be treated with caution.

The obtained temperatures are significantly lower (by few hundreds K) than those predicted 
by a solar metallicity isochrone for the main sequence, which, however, still fits within 
2$\sigma$. Nevertheless, we investigated the possibility that ASAS-052 is more metal rich than 
the Sun. A very good agreement between the model and our measurements was found for 
a very high value of [M/H]=0.5~dex. The formally best-fitting isochrone of this [M/H] 
is for the age of 4.0~Gyr, mostly constrained by the primary's radius. Unfortunately, 
relatively large mass and temperature errors do not allow for 
precise age determination. The system would be slightly younger, provided that [M/H] is lower. 
For instance, the best-fitting solar-composition model was found for the age of 
3.5~Gyr. The eccentricity of the orbit is significant (0.145$\pm$0.005), and the time scale of 
circularisation of the orbit, given in 
Tab.~\ref{tab_par_all}, is 1.56~Gyr, which is less than the age estimated from isochrones. 
The eccentricity may be, however, pumped by the presence of the component B. Our astrometric
solution (see further Sections) suggests that the A+B orbit is highly eccentric, and not
coplanar with the eclipsing orbit.

Notably, parameters of the low-mass secondary match the 4.0~Gyr, [M/H]=0.5 isochrone.
Its radius is nicely reproduced by main-sequence models, which is not always the case for
stars of such mass. The X-ray emission detected from ASAS-052 definitely shows that the system 
contains active stars, but we can't say which ones are active, or even how many. 

To summarize, considering the limitations of our analysis, we refrain ourselves from any 
conclusive statements about the exact age or metallicity of ASAS-052. We can only 
conclude that the system's [M/H] is likely higher than 0, and all components are on 
the main sequence. The 4.0~Gyr [M/H]=0.5 isochrone we adopted reproduces our results 
fairly well, especially the temperatures, but [M/H]=0.0 models fit within errors as well.
We would like to stress that we adopted this unusually high metal content only to
match the obtained temperatures.
It is possible that our colour-based $T_{\rm eff}$ scale was affected by the reddening, 
which we have not taken into account. To shift the colour indices to solar metallicity
values, one would have to assume $E(B-V)$ of the order of 0.1~mag. This seems to be
unlikely at a distance of $\sim$100~pc. To confirm (or disprove) our $T_{\rm eff}$
scale, independent distance and/or reddening evaluation is required.

\subsubsection{The non-eclipsing pair}

\begin{table}
\centering
\caption{Orbital parameters of ASAS-052~B from V2FIT.}\label{tab_par_A052B}
\begin{tabular}{lccc}
\hline \hline
Parameter & Value & $\pm$ \\
\hline
$P$ [d]			& 21.5704	& 0.0005 \\
$T_p$ [JD-2450000]	& 2848.68	& 0.10 \\
$K_1$ [km/s]		& 51.55 	& 0.08 \\
$K_2$ [km/s]		& 56.53	& 0.08 \\
$q$			& 0.912 	& 0.002 \\
$e$			& 0.615	& 0.001 \\
$\omega$ [$^\circ$]	& 100.43	& 0.17 \\
$\gamma_1$ [km/s]	&  13.99	& 0.07 \\
$\gamma_2-\gamma_1$ [km/s] & 0.12 	& 0.08\\
$\mathcal{M}_1 \sin^3{i}$ [M$_\odot$] & 0.714 & 0.003 \\ 
$\mathcal{M}_2 \sin^3{i}$ [M$_\odot$] & 0.661 & 0.003 \\ 
$a \sin{i}$ [R$_\odot$] & 38.67	& 0.05 \\
$rms_{\rm RV,1}$ [km/s]\tablefootmark{a} & \multicolumn{2}{c}{ 0.112/0.139 } \\
$rms_{\rm RV,2}$ [km/s]\tablefootmark{a} & \multicolumn{2}{c}{ 0.209/0.216 } \\
\hline
\end{tabular}
\tablefoot{\tablefoottext{a}{Without and with IRCS.}}
\end{table}

\begin{figure}
\centering
\includegraphics[width=0.95\columnwidth]{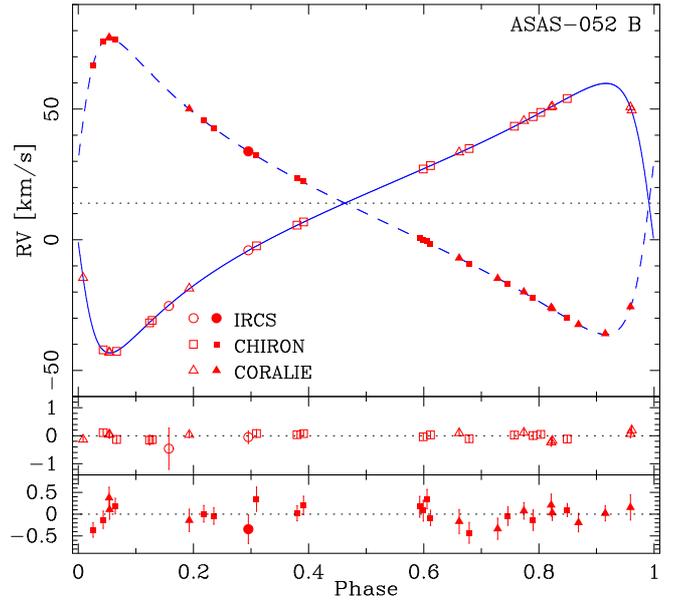}
\caption{Same as Fig.~\ref{fig_rv_A052A}, but for the non-eclipsing pair 
ASAS-052~B. \label{fig_rv_A052B}}
\end{figure}

The orbital parameters of ASAS-052~B, derived from the RV fit, are given in Table~\ref{tab_par_A052B}.
The observed and modelled RVs are shown in Figure~\ref{fig_rv_A052B}. We reach a very
good precision of 0.42\% in $\mathcal{M} \sin^3{i}$ of both components. From the height of
peaks in the CCF, we can deduce that brightness of Ba and Bb is somewhere between the components
of A. Since the system seems to reside on the main sequence, same thing can be said about
masses. Moreover, fractional amounts of the third light ($l_3$) in $I_{\rm C}$ and $R_{\rm C}$ 
(Tab.~\ref{tab_par_all}) suggest that this pair is redder, thus cooler than A. The fact that 
$l_3$ is even more significant in $V$, and larger than the estimates from SOAR observations,  
may be explained by a larger photometric aperture and pixel scale of the ASAS cameras, in 
comparison with MITSuME -- faint background sources were included in the measured flux.

We have checked the residuals of light curve fits, and found no obvious
signs of eclipses with the period of 21.5704~d. This means that the inclination of B's orbit
can not be close to 90$^\circ$. The rigid, conservative limits for the masses are 
therefore <0.662:1.032> and <0.604:0.941>~M$_\odot$ for Ba and Bb, respectively.

We took the 4.0~Gyr isochrone for [M/H]=0.5, and attempted to estimate the true masses by
comparing the observed and predicted absolute magnitudes in $V,R_{\rm C}$, and $I_{\rm C}$. 
We incorporated the distance estimate from JKTEBOP, the adopted $\Delta mag$ between 
A and B from SOAR observations (Tab.~\ref{tab_soar}), and mass ratio from Tab.~\ref{tab_par_A052B}. 
The total absolute magnitudes of ASAS-052~B should be 5.6(3)~mag in $V$,
5.1(3)~mag in $R_{\rm C}$, and 4.6(3)~mag in $I_{\rm C}$ (errors include uncertainties in 
$\Delta mag$, distance, and observed magnitudes of pair A).
We have found that these total magnitudes and the given mass ratio are reproduced by a pair 
of 0.89(4) and 0.81(4)~M$_\odot$ stars. The inclination angle of the inner, 
21.57-day orbit would thus be 69$^\circ$, and no eclipses would occur. 

\subsubsection{Wide orbit of ASAS-052~AB}

\begin{figure}
\centering
\includegraphics[width=0.9\columnwidth]{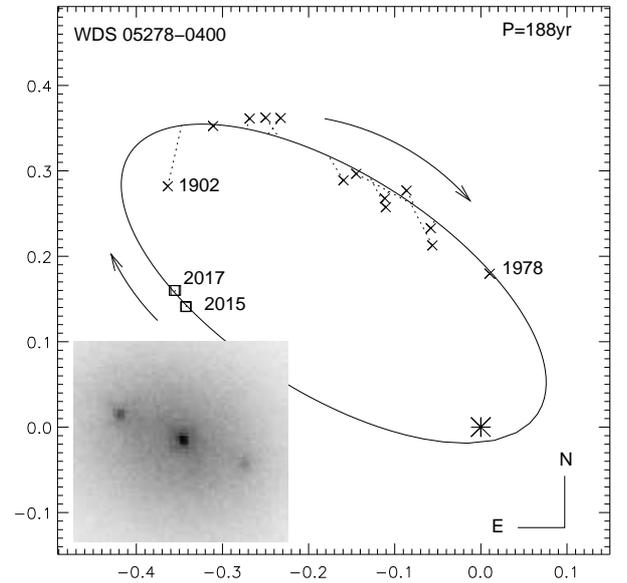}
\caption{Astrometric orbit of the ASAS-052~B relatively to A
(large asterisk symbol). The pre-2015 WDS measurements are shown with crosses, 
and the SOAR data with squares. Dates of some observations are labelled, 
orientation and the direction of orbital motion is shown.
The inset shows a piece of SOAR interferogram, with the same orientation as the orbit.
\label{fig_052wide}}
\end{figure}

\begin{table}
\centering
\caption{Orbital parameters of the ASAS-052~AB system.\label{tab_052ast}}
\begin{tabular}{lcc}
\hline \hline
Parameter & Value & $\pm$ \\
\hline
$P$ [yr]	& 188  & 41 \\
$T_{\rm P}$ [Bess. year]& 1994 & 6 \\
$T_{\rm P}$ [JD-2400000]& 49300 & 2100 \\
$e$		& 0.92 & 0.07 \\
$\hat{a}$ ['']	& 0.44 & 0.17 \\
$\Omega$ [$^\circ$] & 90 & 11 \\
$\omega$ [$^\circ$] & 245 & 15 \\
$i$ [$^\circ$]	& 117 & 14 \\
\hline
\end{tabular}
\end{table}

We have combined the new and old astrometric measurements, and for the first time 
found solution of the system's outer orbit. These orbital parameters are presented 
in Table~\ref{tab_052ast}. The orbit is very eccentric, with the projected angular
separation at pericentre of $\sim$33~mas, which explains the lack of astrometric 
observations after 1978. We used our direct determination of the total mass of A 
(1.638~M$_\odot$), the indirect estimate for B (1.694~M$_\odot$), the period of 
the orbit, and its projected major semi-axis $\hat{a}$ to estimate the dynamical 
parallax. We got 9$\pm$4~mas, or 110$\pm$50~pc. This is in agreement with 
the value calculated by JKTABSDIM. The precision is however much worse, hampered 
mainly by the large relative uncertainty of $\hat{a}$.

The current orbital solution and distance from JKTABSDIM predict that at the physical
separation at pericentre is as small as $\sim$4~AU. We can thus expect strong gravitational
interactions between two pairs, which possibly increase their eccentricities. 
However, for the pair A we can exclude the Mazeh-Shaham mechanism (MSM), 
which produces periodic eccentricity variation known as the Kozai cycles 
\citep{koz62,maz79}. In order to work, Kozai cycles time-scale must be shorter than 
the period of the inner orbit’s pericentre precession. We used the formulae given 
in \citet{fab07} to estimate both quantities. In case of ASAS-052~A relativistic 
precession is the dominant one, and its period is $\sim$20\,000~yr. Meanwhile, the 
predicted Kozai cycle time-scale is about 68\,000~yr, so the main condition for the 
MSM to work is not met.

\subsection{ASAS-065}

\begin{figure}
\centering
\includegraphics[width=0.95\columnwidth]{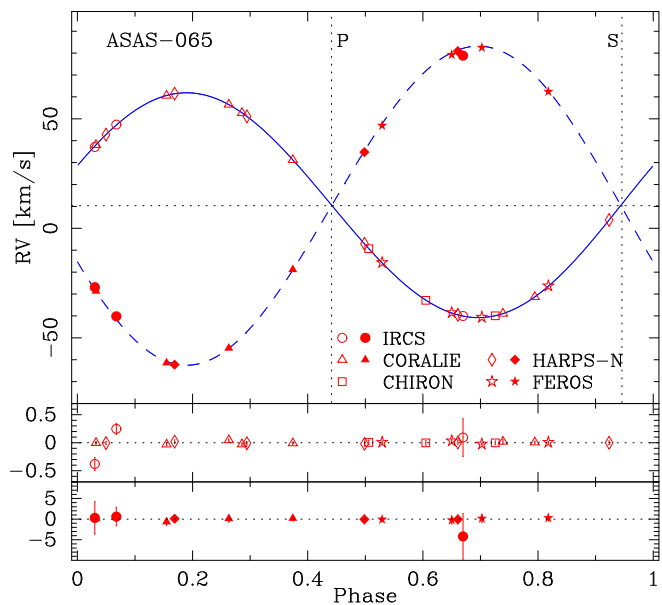}

\caption{Same as Fig.~\ref{fig_rv_A052A}, but for the eclipsing binary
ASAS-065. \label{fig_rv_A065}}
\end{figure}

\begin{figure*}
\centering
\includegraphics[width=0.7\textwidth]{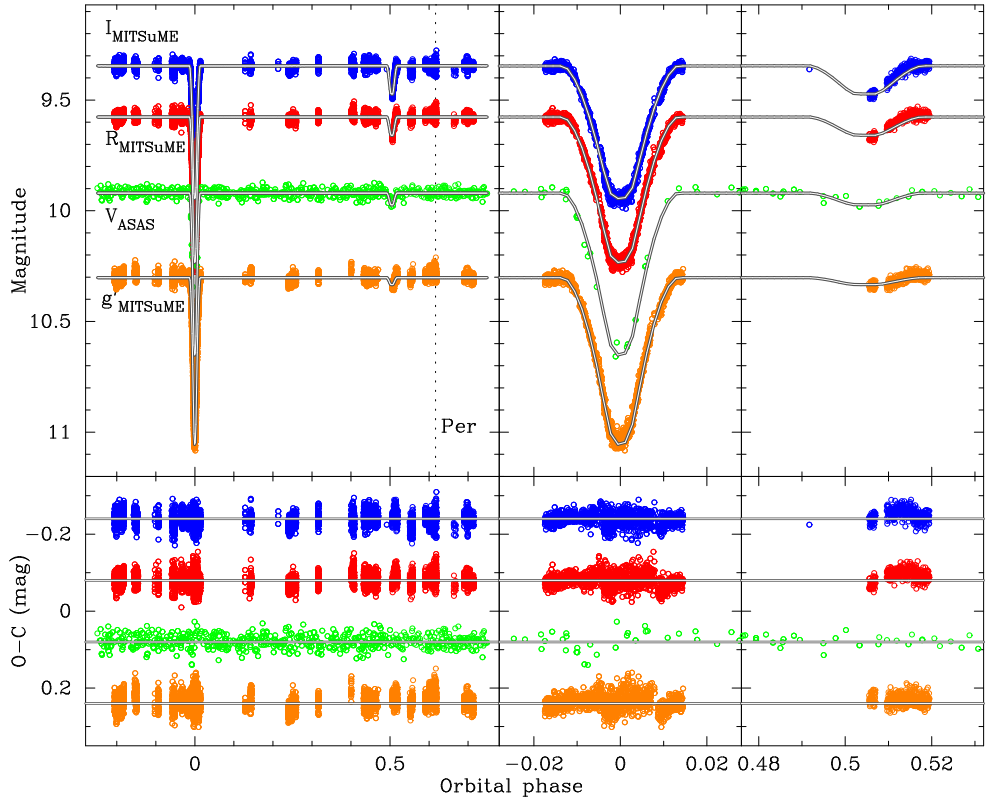}
\caption{Same as Fig.~\ref{fig_lc_A052}, but for ASAS-065, with
the MITSuME $g'$-band curve added. \label{fig_lc_A065}}
\end{figure*}

\begin{figure}
\centering
\includegraphics[width=0.8\columnwidth]{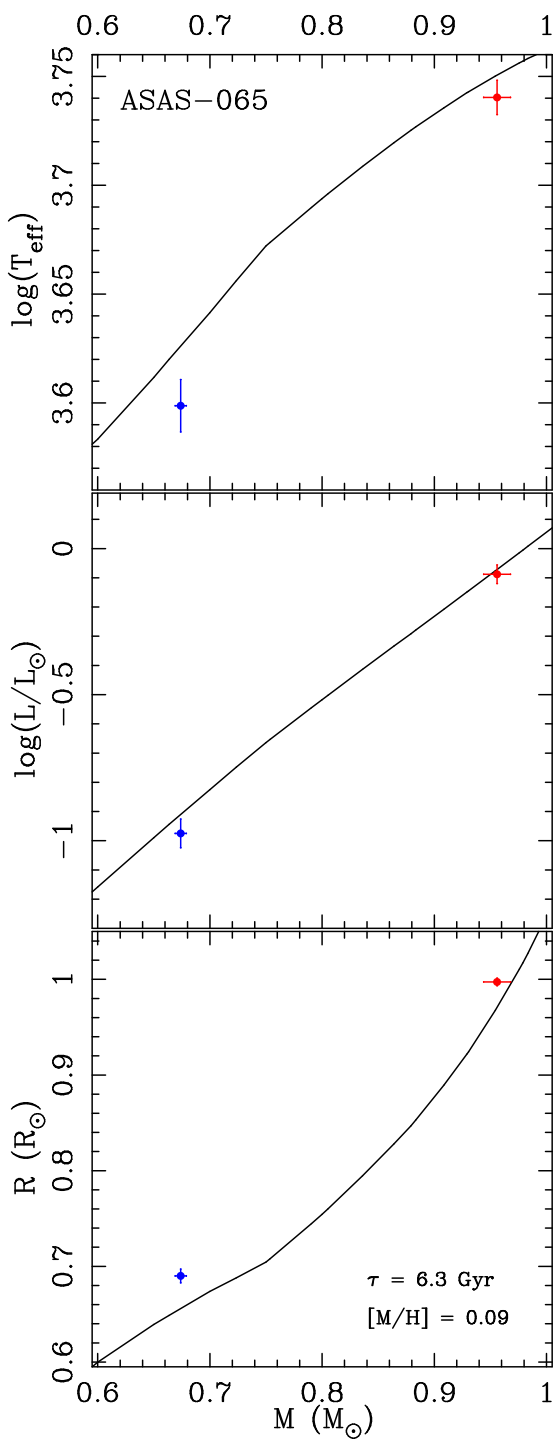} 
\caption{Same as Fig.~\ref{fig_iso_A052}, but for the eclipsing binary
ASAS-065. The isochrone is for 6.3 Gyr and metallicity of +0.09 dex. \label{fig_iso_A065}}
\end{figure}

The results of our analysis of ASAS-065 are presented in Tab.~\ref{tab_par_all}.
Figure~\ref{fig_rv_A065} contains the RV curves, while four light curves are 
shown in Figure~\ref{fig_lc_A065}. This is the system with the best precision in mass 
and radius determination: 1.29+0.71\% uncertainty in mass of the primary and secondary, 
respectively, and 0.40+1.05\% analogously for the radii. This is due to the narrowest lines
(hence highest RV precision), longest orbital period, and good coverage of both minima,
including the totality of the secondary. Worth noting is the fact that the $rms$ of the RV fit 
of the primary from the optical spectra only is 19~m/s, sufficient to detect massive planets
on circumbinary orbits.

The primary's contribution in $V$ is 95\% to the flux of the binary, and 85\% to the total flux.
The source of the third light contamination are possibly two faint stars. One is located about 5 asec
from the DEB, and has entries in several catalogues. It was likely within the photometric
aperture of ASAS. Its distance and proper motion listed in GDR2 suggest that it is 
gravitationally bound to ASAS-065, which makes the system a physical triple. 
The GDR2 gives temperature of $3760^{+860}_{-340}$~K, which makes it cooler than the 
secondary of the eclipsing pair. The other is a visual companion, separated only about 
0.45~asec from ASAS-065, which we have seen on acquisition and guiding images during our 
Subaru observations. Assuming that it is the only contaminant of MITSuME photometry, it appears to 
be bluer than the secondary and brighter in $g'$, but fainter in $R_{\rm C}$ and $I_{\rm C}$, which suggests 
it is a background star. Confirmation should, however, come from dedicated adaptive optics
observations. 

At the beginning it was somewhat uncertain to us how the secondary or the third light(s) affect 
the results of spectroscopic analysis. The most recent one, from RAVE~DR5 \citep{kun17}, gives 
the calibrated $T_\mathrm{eff}=5330(60)$~K, $[M/H]=-0.10(9)$~dex, and $\log(g)=4.43(8)$~dex. The 
agreement with our results is within 2$\sigma$, even though this analysis does not 
take into account binarity, nor third light. Note that the RAVE spectra are taken in the $I$ 
band, where the contribution from the primary to the total flux drops to 79\%, and with much 
lower spectral resolution. This is probably the reason why \citet{min17} underestimated the 
mass of ASAS-065, and obtained 0.832~M$_\odot$, with the 3$\sigma$ uncertainty range 
<0.741:0.933>~M$_\odot$. One can see that it is only marginally in agreement with our dynamical 
mass of the primary (<0.920:0.992>~M$_\odot$ 3$\sigma$ range; Tab.~\ref{tab_par_all}).

We used our estimates of $T_{\rm eff}$ and apparent $V,R_{\rm C},I_{\rm C}$ magnitudes (after correction 
for the third light) to calculate the distance with JKTABSDIM. The nine individual values 
produced by the code gave the weighted average of 108(10)~pc when no extinction was assumed.
This is in good agreement ($\sim$1.4$\sigma$) with GDR2, which gives 92.72(25)~pc.
The GDR2 value can be reached for $E(B-V)\sim0.12$~mag.

Comparison with PARSEC isochrones for [M/H]=0.09~dex gave the estimated age of 6.3~Gyr
(Fig.~\ref{fig_iso_A065}). This age may be considered surprising for the seemingly high metal 
content, but one should note that the orbit is nearly circular and both components seem to rotate 
synchronously, despite the period of $\sim$8.2~d (the circularisation time scale, as estimated 
by JKTABSDIM, is $\sim$19~Gyr), and the metallicity is consistent with solar within the error bars. 
The 6.3~Gyr isochrone matches the primary very well (within 1$\sigma$) on all three planes.
This age suggests that the primary ends its main sequence evolution, and slowly moves towards 
the sub-giant branch. The secondary's luminosity is also reproduced correctly, $T_\mathrm{eff}$
is within 2$\sigma$, but its radius is not. The measured $R$ is larger than the model, 
and $T_\mathrm{eff}$ is clearly lower. This is however a common feature observed
in many low-mass stars in eclipsing binaries. The most popular explanation involves stellar
activity, enhanced by rapid rotation (when the star is tidally locked in a short-period binary)
through some version of a dynamo mechanism. However, several well-measured systems, like
\object{V636~Cen} \citep{cla09}, \object{LSPM~J1112+7626} \citep{irw11}, 
\object{ASAS~J045304-0700.4} \citep{hel11a}, or \object{KOI-126}~BC \citep{fei11} 
seem to contradict this scenario. The ASAS-065 system is an
X-ray source, and we can see clear emission in cores of Ca H and K lines (in HARPS-N spectra).
Therefore, we can securely claim that chromospheric activity in the system is strong, and
may be the reason why the secondary is inflated, but is likely not enhanced by rotation.
ASAS-065 clearly shows narrow spectral lines, the velocity of synchronous rotation of
the secondary (from JKTABSDIM) is only 4.2 km/s. Notably, a similar situation occurs in V530~Ori 
\citep[$P=6.11$~d, $M_1+M_2 = 1.004+0.596$~M$_\odot$; ][]{tor14}, where authors managed 
to reproduce observed radii with models that assume existence of strong magnetic fields 
in both components.

Our results are precise enough to be used for reliable tests of stellar evolution models. 
This system seems to be interesting from this point of view, as it is composed of a solar 
analogue and a low mass star. There are only a few similar, precisely measured DEBs known, e.g. 
\object{V1174~Ori} \citep[1.009+0.731~M$_\odot$;][]{sta04}, IM~Vir \citep[0.981+0.664~M$_\odot$;][]{mor09b}, 
or the aforementioned V530~Ori. The first one does not have its [M/H] estimated. 
Few other systems, like \object{1SWASPJ011351.29+314909.7} 
\citep[0.945+0.186~M$_\odot$;][]{gom14}, or ASAS-052~A show similar characteristics, 
but do not have their parameters measured that well. 
The secondary is also interesting by itself, as it fills a gap in well-measured stars
of masses around 0.65-0.7~M$_\odot$. The DEBCat lists only four such objects, secondaries of
\object{ASAS~J082552-1622.8} \citep[0.687~M$_\odot$, 0.699~R$_\odot$; ][]{hel11a}, 
\object{V404~CMa} \citep[0.662~M$_\odot$, 0.680~R$_\odot$; ][]{roz09}, 
\object{KIC~6131659} \citep[0.685~M$_\odot$, 0.639~R$_\odot$; ][]{bas12}, and
IM~Vir (0.664~M$_\odot$, 0.681~R$_\odot$). One should also mention the primaries of 
AK~For \citep[0.697~M$_\odot$, 0.687~R$_\odot$; ][]{hel14}, and
\object{T-Cyg1-12664} \citep[0.680~M$_\odot$, 0.799~R$_\odot$; ][]{igl17}, which are not in DEBCat 
due to slightly worse precision of measurements. Except KIC~6131659~B, which 
has a long orbital period of 17.53~d, all these stars have inflated radii, 
and are rotating synchronously in tidally-locked configurations, although AK~For~A matches 
the theoretical isochrone within uncertainties ($\sim$2.9\% in radii).

\subsection{ASAS-073}

\begin{figure}
\centering
\includegraphics[width=0.95\columnwidth]{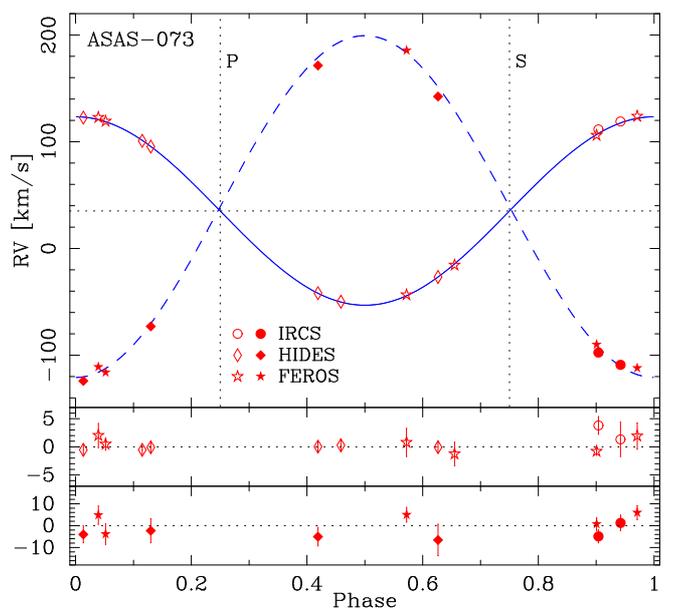}
\caption{Same as Fig.~\ref{fig_rv_A052A}, but for the eclipsing binary
ASAS-073. The orbit of this system is circular, so the zero phase is set 
to the time of quadrature, and eclipses occur exactly in phases 0.25 and 
0.75. \label{fig_rv_A073}}
\end{figure}

\begin{figure*}
\centering
\includegraphics[width=0.7\textwidth]{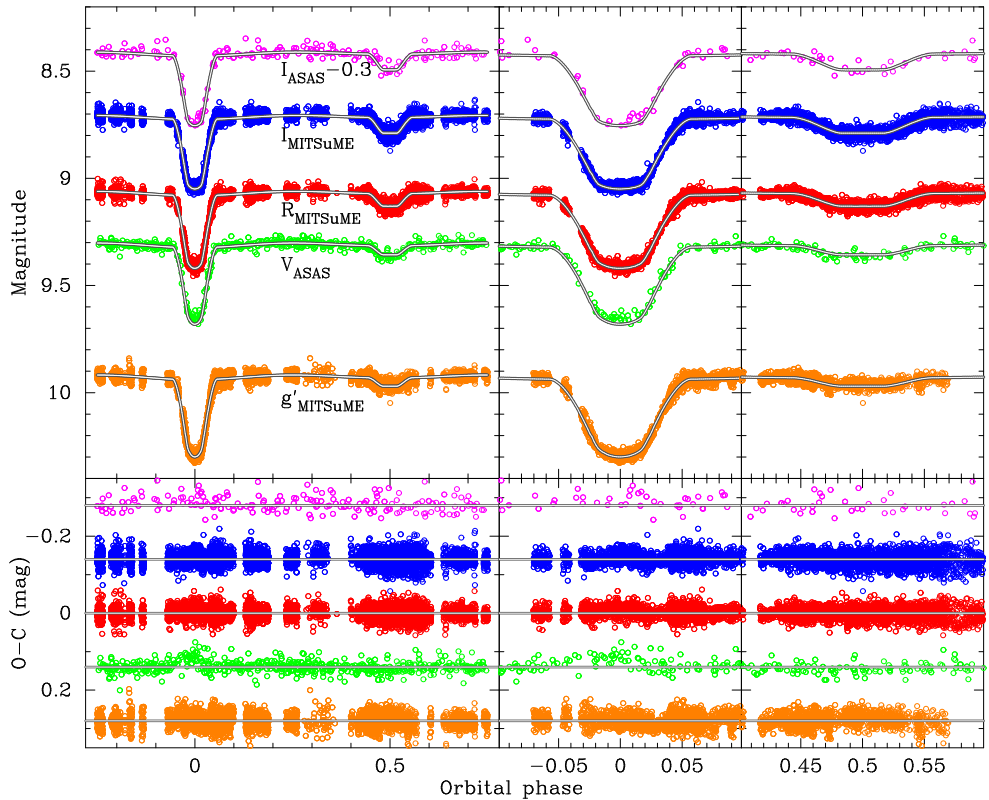}
\caption{Same as Fig.~\ref{fig_lc_A065}, but for ASAS-073, with
the ASAS $I$-band curve added, and shifted vertically for clarity.
Without the shift, it would overlap with $I_{MITSuME}$.
The orbit is circular, so the phase of pericentre passage is 
not defined. \label{fig_lc_A073}}
\end{figure*}

The most massive eclipsing binary in this study is composed of an F-type
primary and a K-type secondary. Figure~\ref{fig_rv_A073} shows the RV curves, 
five light curves are presented in Figure~\ref{fig_lc_A073}, and the resulting parameters
in Tab.~\ref{tab_par_all}. Short period, circular orbit, ellipsoidal variations
of the light curve, and broad spectral lines strongly suggest tidal equilibrium
in this system. We obtained a relatively good precision in masses --
 2.3 and 1.6\% for the primary and secondary, respectively, hampered by rotationally
broadened lines. Our precision in radii determination is even better -- 0.72+1.34\% -- 
thanks to good quality MITSuME light curves. Since no third light had to be assumed, 
for distance determination with JKTEBOP we could use observed total magnitudes in more 
bands that just the ones we observed in. We used the $T_\mathrm{eff,1}$ from 
spectral analysis, and $T_\mathrm{eff,2}$ calculated using the temperature 
fraction derived with PHOEBE. Our adopted distance -- 170(5)~pc -- assumes no 
extinction, is an average of 16 single values, and their $rms$ is taken as the 
uncertainty. It is in a very good agreement with 
{\it Gaia}~DR2, which gives 171(2)~pc. 

\begin{figure}
\centering
\includegraphics[width=0.8\columnwidth]{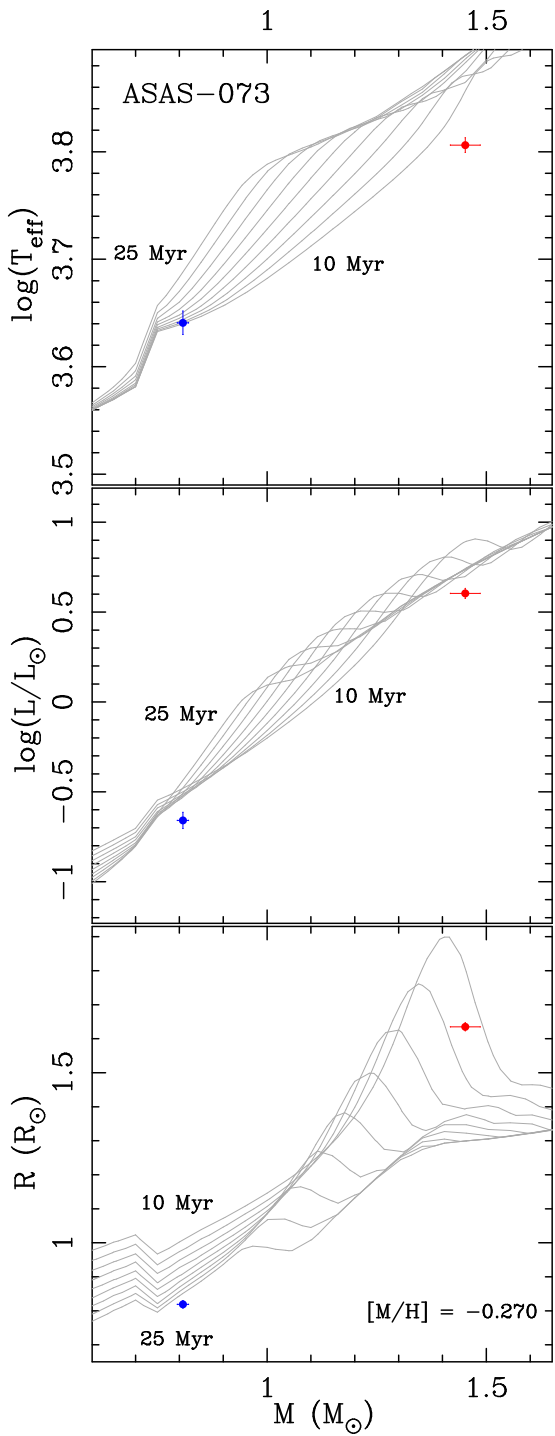} 
\caption{Same as Fig.~\ref{fig_iso_A052}, but for the eclipsing binary
ASAS-073, and a family of [M/H]=$-0.27$ isochrones, spanning from 
$\log(\tau)=7.0$ (10~Myr) to $\log(\tau)=7.4$ (25~Myr), every 
$\Delta\log(\tau) = 0.05$ (grey lines). \label{fig_iso_A073}}
\end{figure}

This agreement suggests that our temperature scale is correct, but it causes problems when
comes to comparison with isochrones, which is shown in Figure~\ref{fig_iso_A073}.
For the given masses and our metallicity of $-0.27$~dex, one would expect temperatures at the main 
sequence around 7800~K for the primary and 5400~K for the secondary. Meanwhile, our 
values are 1000-1400~K lower. Also the available literature estimates \citep[i.e.][]{amm06,gai18} 
give values around 6300~K. The total secondary eclipse, and lack of third light, allow 
for direct determination of observed primary's colours. From $(V-I_{\rm C})=0.56(2)$~mag, 
$(R_{\rm C}-I_{\rm C}) = 0.34(2)$~mag, $(V-R_{\rm C})=0.23(2)$~mag, and calibrations of \citet{wor11}, 
we get $T_{\rm eff,1} = 6240(590)$~K. Similar procedure for the secondary gives 
4400(150)~K. Even though the uncertainties are large, these values are well below the ones
expected at the main sequence, and in agreement with those in Table~\ref{tab_par_all}.

Our results can be, at least qualitatively, explained by young age of the system. 
Proprties of the primary and secondary are best reproduced (within 3$\sigma$) 
by 10 and 25~Myr isochrones, respectively (assuming [M/H]=$-0.27$~dex). This means the
pre-main sequence (PMS) phase. The isochrones in Fig.~\ref{fig_iso_A073} are shown 
for ages 10-25~Myr. The age in PARSEC is counted from a position on the Hayashi
track, and from a model that is artificially set to have the central temperature of
$\sim$10$^5$~K \citep{bre12}. It is uncertain if two protostars of substantially different
masses, forming from the same molecular cloud, reached the Hayashi track
and this particular temperature at the same moment.
If not, a difference in age of $\sim$15~Myr can be explained. Therefore, we suspect that 
ASAS-073 may be a rare case of a PMS eclipsing binary. It is not in disagreement with the theory of 
tidal interactions, which predicts circularisation of the orbit of ASAS-073 after $\sim$1.5~Myr. 
We would also like to note that in the UV images from the GALEX satellite, at the position of
ASAS-073 there is a strong, point-like source that seems to be surrounded by a fainter 
extended emission. No X-ray source is related to the target, and no significant emission
is seen in cores of Ca~H and K lines in the spectra.

The primary's radius and effective temperature are formally reproduced by ``older'' 
isochrones of ages 1.4 and 2.5~Gyr, respectively. Both would require the secondary to be
$>$10\% smaller. Discrepancies in radius occur for 0.8~M$_\odot$ stars, but usually are not
that large.
Reaching the temperatures predicted at the main-sequence phase and the {\it Gaia} 
distance simultaneously requires a substantial amount of interstellar extinction 
and reddening, around $E(B-V)\sim0.2-0.25$~mag. Independent comparison with PARSEC
isochrones on the mass vs. $(V-I_{\rm C})$ plane clearly shows both components to be $\sim$0.3-0.4~mag
too red for the MS. The $E(B-V)$ given by the Galactic Dust Reddening and Extinction on-line 
service\footnote{\tt http://irsa.ipac.caltech.edu/applications/DUST/} is 
0.194(6), which is close to the required value, but would mean that
all, or at least majority of the dust in the line of sight towards ASAS-073 is
located closer than 170~pc, which we find unlikely,considering the target
lays in the Galactic disk ($b=+5.415^\circ$). Also, we measured the equivalent
width (EW) of the interstellar sodium D1 line in FEROS and HIDES spectra, and
found it to be 0.025(5)~\AA. According to calibrations by \citet[][see also 
references therein]{mun97}, it should be about 20 times larger to explain 
$E(B-V)\sim0.2$~mag. 

Furthermore, the observed values of $l_2/l_1$ are in strong disagreement
with those predicted by 1.4 and 2.5 Gyr isochrones, but can be reproduced by the 
10-25~Myr isochrones. From the fact that the secondary eclipse is total and well sampled, 
the flux ratios are robust and very well constrained in all bands, especially in the 
MITSuME data, and independent on other parameters, like radii or effective temperatures, 
or the photometric calibrations (see Sect.~\ref{sec_mits_photo}). Their inconsistencies 
with ``older'' isochrones further supports the PMS scenario. 

Finally, we checked the Galactic kinematics of ASAS-073. 
Using the parallax and proper motion from {\it Gaia}~DR2 and our value of the systemic 
velocity $\gamma$, we have calculated the spatial motion components: $U=37.3(3)$, $V=-15.4(3)$,
and $W=-18.3(2)$~km/s (no correction for the solar movement has been done). These 
values put ASAS-073 within the thin disk, at the edge of space occupied by a moving 
group called Coma Berenices or ``local'' \citep{fam05,sea07}. \citet{fam05} have shown 
that ages of stars from this group vary from several to few hundreds of Myr.
This membership, if confirmed, would also support the ``young'' isochrone age of 10-25~Myr, 
and make the main sequence stage less probable.

We therefore conclude that ASAS-073 is probably a PMS eclipsing system, and if so,
it is valuable for testing models at early stages of stellar evolution. There are not 
many PMS DEBs known, especially with component masses below 1.5~M$_\odot$. 
\citet{cor15} listed seven such systems (see references therein) and presented one new 
candidate. Only two of those binaries are currently in the DEBCat: \object{ASAS~J052821+0338.5}
\citep{ste08} and V1174~Ori \citep{sta04}. More recently, four more cases have been
reported by \citet{kra15}, \citet[][one probably composed of two brown dwarfs]{dav16}, and
\citet{lac16}. The PMS eclipsing
binaries are heavily under-studied, and every new example is highly valuable. The definitive 
confirmation of the evolutionary state of ASAS-073 requires better knowledge of 
the temperatures and extinction towards the target. Notably, this system is brighter 
than majority of the PMS examples known to date, which 
makes it a relatively easy target for further studies, which are highly encouraged.

\section{Summary}
In this study we present physical parameters of three targets from our 
large program aimed for characterisation of interesting eclipsing binary systems.
In order to quickly but securely measure RVs of secondary components of 
the presented cases we decided to observe in the IR. This also allowed us to
confirm if some RV measurements from the optical spectra are real, and have not
been confused with sidelobes of the CCF or other artefacts. 
All eclipsing pairs show low mass ratio, which makes fitting an isochrone
more difficult than in case of similar-mass stars. All three examples are, however, 
very different and interesting in their own way. ASAS-052 is a quadruple-lined (triple
in visible light), SB2+SB2 multiple with eclipsing pair, and all four stellar 
masses estimated. RV measurements of the faintest component were only made in the IR. 
The astrometric orbit has been determined for the first time, and suggests
strong dynamical interactions between pairs Aa+Ab (eclipsing) and Ba+Bb 
(non-eclipsing). Both ASAS-052~A and ASAS-065 are composed of a solar-analogue primary 
and a low-mass secondary, but the latter has its parameters measured much more precisely, 
sufficiently for testing stellar evolution models, and is actually composed of at 
least three stars, including a distant, low-mass companion. It also allows for very 
precise RV measurements, at the level adequate to search for massive circumbinary planets. 
Moreover, its significant activity at a relatively old age, not enhanced by
rotation, is somewhat unusual. Finally, ASAS-073 is probably a new late-type PMS system, 
one of the brightest known to date, which gives great opportunity for further studies.
 
Our study shows the advantage of infra-red spectroscopy in observations of low-mass
stars. We managed to directly calculate RVs of the faint, cool components 
with a relatively small use of telescope time. We were able to quickly obtain
useful data, crucial for modelling and understanding the nature of the studied systems.
There are still many open questions regarding the low-mass stars, and their image will
not be complete without proper attention drawn to high-contrast systems. Since the
number of well-studied cases is very low, it is still unclear if, for example, the
K and M-type stars paired with F or G-type primaries show systematic differences
with respect to the double-M and double-K binaries. There are potentially many bright 
DEBs that appear to be single-lined in optical, but may occur to be SB2 in IR. 
They can be identified as, for example, a by-product of transit search surveys 
\citep{bea07,tri13,tri17}. The growing number of high-stability IR spectrographs, 
like the IRD at Subaru \citep{kot14} or GIANO at TNG \citep{ori14}, will make 
these stars easier to study in more details, which they truly deserve.

\begin{acknowledgements}
We would like to thank the anonymous Referee for valuable comments and suggestions that
helped us to improve the work. We would also like to thank the staff of the 
ESO La Silla, Geneva, and Cerro Tololo observatories, as well as SOAR, TNG, and Subaru 
telescopes for their support during observations. 
We also wish to recognize and acknowledge the very significant cultural role and 
reverence that the summit of Maunakea has always had within the indigenous Hawaiian
community. We are most fortunate to have the opportunity to conduct observations from this mountain.

\\ This publication is based on data collected: 
at the Subaru Telescope, which is operated by the National Astronomical Observatory of Japan; 
through CNTAC proposals CN-2012B-36, CN-2013A-93, CN-2013B-22, CN-2014A-44, and CN-2014B-67;
at the European Southern Observatory, Chile under programmes 088.D-0080, 090.D-0061, 091.D-0145; 
at the Southern Astrophysical Research (SOAR) telescope, which is a joint project of the 
Minist\'{e}rio da Ci\^{e}ncia, Tecnologia, e Inova\c{c}\~{a}o (MCTI) da Rep\'{u}blica Federativa do Brasil, the U.S. National Optical Astronomy Observatory (NOAO), the University of North Carolina at Chapel Hill (UNC), and Michigan State University (MSU);
with the HARPS-N spectrograph on the 3.58 m Italian Telescopio Nazionale Galileo (TNG) operated on the island of La Palma by the Fundaci\'{o}n Galileo Galilei of the INAF (Instituto Nazionale di Astrofisica) at the Spanish Observatorio del Roque de los Muchachos of the Instituto de Astrofisica de Canarias (programme OPT14B~45 from OPTICON common time allocation process for EC supported trans-national access to European telescopes). 

\\ This research has made use of the Washington Double Star Catalog maintained at the U.S. Naval Observatory.
This work has made use of data from the European Space Agency (ESA) mission {\it Gaia} (\url{https://www.cosmos.esa.int/gaia}), processed by the {\it Gaia} Data Processing and Analysis Consortium (DPAC, \url{https://www.cosmos.esa.int/web/gaia/dpac/consortium}). Funding~for the DPAC has been provided by national institutions, in particular the institutions participating in the {\it Gaia} Multilateral Agreement. 

\\ KGH, EN, and MR acknowledge support provided by the Polish National Science Center through grants no. 2016/21/B/ST9/01613, 2014/13/B/ST9/00902, and 2015/16/S/ST9/00461, respectively. 
This work was partially supported by JSPS KAKENHI Grant Number 16H01106.

\end{acknowledgements}


\begin{appendix}
\section{Individual RV measurements}
In Tables~\ref{tab_rv_052} to \ref{tab_rv_073} we list all the RVs used in this study. 
We also give the final measurement errors $\epsilon$ and residuals of the fit $(O-C)$, together
with the exposure time $t_\mathrm{exp}$ of a given spectrum, and its signal-to-noise ratio SNR 
calculated at around 5500~\AA\,for optical spectra and 12900~\AA\,for IR. 
When no measurement is given, lines were either not detected,
or blended with another component. The last column shows the telescope/spectrograph used, 
coded as follows: 
CC~=~CTIO 1.5m/CHIRON, 
EC~=~Euler/CORALIE, 
MF~=~MPG-2.2m/FEROS, 
OH~=~OAO-188/HIDES, 
SI~=~Subaru/IRCS,
TH~=~TNG/HARPS-N. 

\begin{table*}
\centering
\caption{Individual RV measurements of ASAS-052 A and B used in this work.\label{tab_rv_052}}
\begin{tabular}{lrrrrrrrrc}
\hline \hline
HJD & $v_1$ & $\epsilon_1$ & $(O-C)_1$ & $v_2$ & $\epsilon_2$ & $(O-C)_2$ & $t_\mathrm{exp}$ & SNR & Tel./Sp.\\
-2450000 & [km/s] & [km/s] & [km/s] & [km/s] & [km/s] & [km/s] & [s] &  & \\
\hline
\multicolumn{10}{l}{\bf ASAS-052 A}\\
6269.730419 &   62.068 & 0.101 & -0.021 &    ---   &  ---  &   ---  &  780 &  40 & CC \\
6314.591631 &  -30.253 & 0.141 &  0.097 &    ---   &  ---  &   ---  &  780 &  40 & CC \\
6338.557052 &   71.557 & 0.115 & -0.024 &    ---   &  ---  &   ---  &  780 &  40 & CC \\
6373.546403 &  -14.165 & 0.112 &  0.161 &    ---   &  ---  &   ---  &  780 &  35 & CC \\
6543.897710 &   59.534 & 0.109 & -0.080 &    ---   &  ---  &   ---  &  900 &  35 & CC \\
6618.626044 &   73.887 & 0.107 &  0.061 &    ---   &  ---  &   ---  &  500 &  35 & EC \\
6696.652663 &   51.515 & 0.125 &  0.013 &    ---   &  ---  &   ---  &  900 &  40 & CC \\
6727.539302 &   14.152 & 0.111 &  0.026 &    ---   &  ---  &   ---  &  780 &  50 & EC \\
6732.528275 &    0.775 & 0.115 & -0.156 &    ---   &  ---  &   ---  &  780 &  25 & EC \\
6734.742599 &    ---   &  ---  &   ---  &  -69.575 & 3.718 &  0.612 & 1800 & 120 & SI \\
6737.720831 &   -2.656 & 0.148 & -0.033 &   53.333 & 2.623 & -1.660 &  760 & 190 & SI \\
6744.549535 &   69.236 & 0.091 & -0.139 &    ---   &  ---  &   ---  &  900 &  60 & CC \\
6753.481577 &   -5.312 & 0.094 & -0.167 &    ---   &  ---  &   ---  &  900 &  55 & CC \\
6939.757371 &   72.105 & 0.120 &  0.148 &    ---   &  ---  &   ---  &  780 &  55 & EC \\
6943.806573 &   16.588 & 0.139 &  0.047 &    ---   &  ---  &   ---  &  900 &  60 & CC \\
6967.745832 &    0.689 & 0.117 & -0.053 &    ---   &  ---  &   ---  &  900 &  25 & EC \\
6969.782964 &   -1.554 & 0.112 & -0.025 &    ---   &  ---  &   ---  &  780 &  35 & EC \\
6976.827734 &   75.301 & 0.109 &  0.003 &    ---   &  ---  &   ---  &  900 &  70 & CC \\
7000.993600 &  -17.772 & 0.314 &  0.259 &   82.308 & 1.829 &  0.984 & 1080 & 180 & SI \\
7001.056749 &  -16.012 & 0.285 & -0.091 &   77.096 & 2.527 & -0.621 & 1080 & 190 & SI \\
7012.704614 &   33.796 & 0.092 &  0.216 &    ---   &  ---  &   ---  &  900 &  55 & CC \\

\hline
\multicolumn{10}{l}{\bf ASAS-052 B}\\
6269.730419 &   27.053 & 0.096 & -0.040 &   -0.183 & 0.239 &  0.080 &  780 &  40 & CC \\ 
6304.647093 &    ---   &  ---  &   ---  &   45.659 & 0.191 &  0.010 &  780 &  35 & CC \\ 
6314.591631 &   34.884 & 0.111 & -0.106 &   -9.348 & 0.238 & -0.427 &  780 &  40 & CC \\ 
6338.557052 &   47.105 & 0.139 &  0.007 &  -22.334 & 0.239 & -0.137 &  780 &  40 & CC \\ 
6344.519993 &  -42.709 & 0.139 & -0.133 &    ---   &  ---  &   ---  &  780 &  45 & CC \\ 
6539.894258 &  -31.805 & 0.179 & -0.154 &    ---   &  ---  &   ---  &  900 &  25 & CC \\ 
6543.897710 &   -2.346 & 0.133 &  0.080 &   32.449 & 0.281 &  0.346 &  900 &  35 & CC \\ 
6563.868932 &    ---   &  ---  &   ---  &   42.757 & 0.191 & -0.038 &  900 &  35 & CC \\ 
6571.857717 &    ---   &  ---  &   ---  &   -0.614 & 0.221 &  0.346 &  900 &  30 & CC \\ 
6574.886786 &    ---   &  ---  &   ---  &  -16.783 & 0.215 & -0.043 &  900 &  30 & CC \\ 
6617.643685 &    ---   &  ---  &   ---  &  -14.798 & 0.241 & -0.335 &  500 &  35 & EC \\ 
6618.626044 &   45.484 & 0.109 &  0.109 &  -19.939 & 0.189 &  0.077 &  500 &  35 & EC \\ 
6619.647581 &   50.708 & 0.143 & -0.233 &  -25.909 & 0.253 &  0.209 &  500 &  30 & EC \\ 
6689.599723 &    ---   &  ---  &   ---  &   76.567 & 0.181 &  0.185 &  900 &  50 & CC \\ 
6696.652663 &    6.793 & 0.090 &  0.078 &   22.277 & 0.213 &  0.197 &  900 &  40 & CC \\ 
6712.554377 &  -30.864 & 0.131 & -0.138 &    ---   &  ---  &   ---  &  900 &  70 & CC \\ 
6722.576279 &    ---   &  ---  &   ---  &    0.558 & 0.239 &  0.171 &  900 &  50 & CC \\ 
6727.539302 &   50.985 & 0.123 & -0.177 &  -26.328 & 0.175 &  0.032 &  780 &  50 & EC \\ 
6728.523318 &    ---   &  ---  &   ---  &  -32.410 & 0.213 & -0.193 &  635 &  30 & EC \\ 
6729.526278 &    ---   &  ---  &   ---  &  -35.955 & 0.177 &  0.020 &  635 &  30 & EC \\ 
6730.518891 &   49.485 & 0.149 &  0.198 &    ---   &  ---  &   ---  &  780 &  30 & EC \\ 
6731.535889 &  -14.547 & 0.092 & -0.133 &    ---   &  ---  &   ---  &  780 &  35 & EC \\ 
6732.528275 &  -43.110 & 0.139 &  0.051 &   77.176 & 0.223 &  0.107 &  780 &  25 & EC \\ 
6734.747693 &  -25.379 & 0.738 & -0.462 &    ---   &  ---  &   ---  & 1800 & 120 & SI \\ 
6737.725690 &   -4.075 & 0.225 & -0.051 &   33.802 & 0.334 & -0.345 &  760 & 200 & SI \\ 
6744.549535 &   28.339 & 0.109 &  0.031 &   -1.681 & 0.173 & -0.086 &  900 &  60 & CC \\ 
6753.481577 &    ---   &  ---  &   ---  &   66.762 & 0.167 & -0.362 &  900 &  55 & CC \\ 
6939.757371 &   33.454 & 0.096 &  0.097 &   -7.008 & 0.277 & -0.168 &  780 &  55 & EC \\ 
6941.826974 &   43.403 & 0.103 &  0.026 &    ---   &  ---  &   ---  &  900 &  65 & CC \\ 
6942.815109 &   48.713 & 0.094 &  0.050 &    ---   &  ---  &   ---  &  900 &  40 & CC \\ 
6943.806573 &   54.027 & 0.119 & -0.118 &  -29.828 & 0.151 &  0.094 &  900 &  60 & CC \\ 
6967.745832 &   50.731 & 0.145 &  0.069 &  -25.662 & 0.291 &  0.155 &  900 &  25 & EC \\ 
6969.782964 &  -43.115 & 0.119 &  0.036 &   77.432 & 0.243 &  0.373 &  780 &  35 & EC \\ 
6972.785254 &  -18.632 & 0.096 &  0.034 &   50.059 & 0.259 & -0.143 &  900 &  50 & EC \\ 
6976.827734 &    5.574 & 0.107 &  0.036 &   23.391 & 0.175 &  0.020 &  900 &  70 & CC \\ 
7012.704614 &  -42.160 & 0.094 &  0.109 &   75.671 & 0.199 & -0.127 &  900 &  55 & CC \\ 
\hline
\end{tabular}
\end{table*}

\begin{table*}
\centering
\caption{Individual RV measurements of ASAS-065 used in this work.\label{tab_rv_065}}
\begin{tabular}{lrrrrrrrrc}
\hline \hline
HJD & $v_1$ & $\epsilon_1$ & $(O-C)_1$ & $v_2$ & $\epsilon_2$ & $(O-C)_2$ & $t_\mathrm{exp}$ & SNR & Tel./Sp.\\
-2450000 & [km/s] & [km/s] & [km/s] & [km/s] & [km/s] & [km/s] & [s] &  & \\
\hline
6381.605521 &  -40.639 & 0.017 & -0.024 &   82.497 & 0.894 &  0.124 &  450 &  55 & MF \\
6382.553993 &  -26.338 & 0.014 &  0.005 &   62.364 & 0.712 &  0.259 &  600 & 120 & MF \\
6519.914854 &  -15.684 & 0.019 &  0.008 &   46.874 & 0.546 & -0.106 &  480 &  55 & MF \\
6520.908621 &  -38.538 & 0.028 &  0.032 &   79.207 & 0.918 & -0.261 &  720 &  80 & MF \\
6560.821911 &   -9.305 & 0.040 &  0.005 &    ---   &  ---  &   ---  & 1020 &  30 & CC \\
6569.856148 &  -32.847 & 0.039 & -0.002 &    ---   &  ---  &   ---  & 1020 &  25 & CC \\
6570.847931 &  -39.936 & 0.032 & -0.002 &    ---   &  ---  &   ---  & 1020 &  25 & CC \\
6634.736878 &   -7.055 & 0.035 & -0.019 &   34.805 & 1.078 & -0.064 & 1800 &  80 & TH \\
6727.585279 &  -31.296 & 0.022 &  0.007 &    ---   &  ---  &   ---  & 1200 &  35 & EC \\
6729.540054 &   38.115 & 0.024 & -0.002 &  -28.676 & 0.768 &  0.117 & 1200 &  45 & EC \\
6730.547507 &   60.625 & 0.026 & -0.028 &  -61.399 & 0.890 & -0.644 & 1200 &  40 & EC \\
6731.623765 &   52.706 & 0.023 & -0.027 &    ---   &  ---  &   ---  & 1200 &  25 & EC \\
6734.780149 &  -40.018 & 0.337 &  0.093 &   78.821 & 5.510 & -4.185 & 1200 &  35 & SI \\
6737.743050 &   37.129 & 0.118 & -0.370 &  -26.884 & 3.936 &  0.287 &  720 & 175 & SI \\
6759.364895 &  -39.489 & 0.030 &  0.006 &   80.897 & 1.098 & -0.063 & 1500 &  40 & TH \\
6940.840189 &  -38.906 & 0.043 &  0.020 &    ---   &  ---  &   ---  & 1200 &  15 & EC \\
6970.719553 &   31.244 & 0.023 & -0.007 &  -18.859 & 0.648 &  0.193 &  900 &  25 & EC \\
7001.077783 &   47.345 & 0.097 &  0.248 &  -40.169 & 2.250 &  0.604 &  800 & 240 & SI \\
7057.430271 &    3.803 & 0.031 &  0.001 &    ---   &  ---  &   ---  & 1800 &  70 & TH \\
7058.467540 &   42.775 & 0.030 & -0.006 &    ---   &  ---  &   ---  & 1800 &  55 & TH \\
7059.447564 &   61.492 & 0.030 &  0.021 &  -62.245 & 0.892 &  0.085 & 1800 &  80 & TH \\
7060.479333 &   51.118 & 0.030 & -0.009 &    ---   &  ---  &   ---  & 1800 &  80 & TH \\
7109.538911 &   56.515 & 0.023 &  0.047 &  -54.710 & 0.848 &  0.111 & 1200 &  30 & EC \\
\hline
\end{tabular}
\end{table*}

\begin{table*}
\centering
\caption{Individual RV measurements of ASAS-073 used in this work.\label{tab_rv_073}}
\begin{tabular}{lrrrrrrrrc}
\hline \hline
HJD & $v_1$ & $\epsilon_1$ & $(O-C)_1$ & $v_2$ & $\epsilon_2$ & $(O-C)_2$ & $t_\mathrm{exp}$ & SNR & Tel./Sp.\\
-2450000 & [km/s] & [km/s] & [km/s] & [km/s] & [km/s] & [km/s] & [s] &  & \\
\hline
5876.752337 &  -43.471 & 2.504 &  0.785 &  185.486 & 3.366 &  4.954 &  450 &  45 & MF \\
5876.872569 &  -15.663 & 2.094 & -1.255 &    ---   &  ---  &   ---  &  600 & 110 & MF \\
5878.775247 &  123.814 & 2.333 &  1.921 & -112.120 & 3.201 &  5.923 &  600 &  90 & MF \\
5878.874412 &  122.686 & 2.135 &  2.010 & -111.070 & 4.184 &  4.791 &  750 & 120 & MF \\
5962.556412 &  105.916 & 0.689 & -0.795 &  -90.091 & 2.912 &  0.709 &  600 &  95 & MF \\
6428.468800 &  119.192 & 1.067 &  0.475 & -116.241 & 4.709 & -3.897 &  420 &  90 & MF \\
6737.808176 &  118.920 & 3.055 &  1.340 & -109.053 & 3.520 &  1.251 &  480 & 230 & SI \\
7000.971042 &  111.538 & 1.578 &  3.806 &  -97.543 & 3.002 & -4.910 &  800 & 230 & SI \\
7024.269438 &  122.535 & 1.038 & -0.526 & -124.139 & 3.846 & -3.999 & 2400 &  65 & OH \\
7059.126708 &  100.739 & 0.886 & -0.533 &    ---   &  ---  &   ---  &  900 &  60 & OH \\
7061.069777 &  -49.913 & 1.127 &  0.279 &    ---   &  ---  &   ---  &  600 &  40 & OH \\
7062.040789 &   95.440 & 0.929 & -0.073 &  -73.021 & 5.546 & -2.321 &  900 &  60 & OH \\
7111.931602 &  -26.746 & 0.831 & -0.052 &  142.357 & 7.197 & -6.572 & 1200 & 100 & OH \\
7755.214215 &  -41.954 & 0.834 &  0.031 &  171.408 & 4.240 & -5.037 & 1200 &  60 & OH \\
\hline
\end{tabular}
\end{table*}

\end{appendix}

\listofobjects

\end{document}